\documentclass[a4paper,10pt,
twocolumn,colorlinks,urlcolor=black,linkcolor=black,citecolor=black,
]{article}
\usepackage{amsmath,amssymb}
\usepackage[dvips]{color}
\usepackage{cite}
\usepackage{hangcaption}
\usepackage{amsmath,amssymb}
\usepackage[dvips]{color}
\usepackage{cite}
\usepackage{eepic}
\usepackage{epic}
\usepackage{hangcaption}
\usepackage{amsmath}
\usepackage{amssymb}
\usepackage{amscd}
\usepackage{amsthm}
\usepackage{multicol}
\usepackage[dvipdfmx,%
 bookmarks=true,%
 bookmarksnumbered=true,%
 colorlinks=true,%
 setpagesize=false,%
 pdfkeywords={TeX; dvipdfmx; hyperref; color;}]{hyperref}
\def\cases{\left\{\begin{array}{ll}}
\def\endcases{\end{array}\right.}
\def\roman{\rm}
\def\bigtimes{\mathop{\mbox{\Large $\times$}}}
\begin{document}
\setcounter{page}{1}
\twocolumn[
\vskip1.5cm
\begin{center}
{\Large \bf 
What is Statistics?
;
The Answer by Quantum Language
}
\vskip0.5cm
{\rm
\large
Shiro Ishikawa
}
\\
\vskip0.2cm
\rm
\it
Department of Mathematics, Faculty of Science and Technology,
Keio University,
\\ 
3-14-1, Hiyoshi, Kouhoku-kuYokohama, Japan.
E-mail:
ishikawa@math.keio.ac.jp
\end{center}
\par
\rm
\vskip0.3cm
\par
\noindent
{\bf Abstract}
\normalsize
\vskip0.5cm
\par
\noindent
Since the problem:
"What is statistics?"
is most fundamental in sceince,
in order to solve this problem,
there is every reason to believe that
we have to
start from the proposal of a worldview.
Recently we proposed measurement theory (i.e.,
quantum language,
or
the linguistic interpretation of quantum mechanics),
which is characterized as the linguistic turn of 
the Copenhagen interpretation of quantum mechanics.
This turn from physics to language does not only extend quantum theory to classical theory but also yield the quantum mechanical world view
(i.e.,
the (quantum) linguistic world view,
and thus, a form of quantum thinking,
in other words, quantum philosophy
). 
Thus, we believe that the quantum lingistic formulation of statistics
gives an answer to
the question:
"What is statistics?".
In this paper,
this will be done through the studies of
inference interval,
statistical hypothesis testing,
Fisher maximum likelihood method,
Bayes method
and
regression analysis
in meaurement theory.

\vskip2.0cm
]

\par

\def\Cal{\cal}
\def\bigstimes{\text{\large $\: \boxtimes \,$}}

\par
\noindent

\par
\noindent
\section{\large Introduction}
\par
\subsection{\normalsize[Animistic worldview] 
$\rightarrow$
[Mechanical worldview] 
$\rightarrow$
[Linguistic worldview]
}
\par
Although research of the worldview
(or, world-description)
has 3000 years or more of history, 
it was always the central theme in science.
The leap (paradigm shift) from the animistic worldview
(i.e.,
\\

\par
\par
\vskip0.5cm
\par
\noindent
\begin{picture}(500,150)
\put(10,70){
{
\put(0,-3){
${\fbox{\shortstack[l]{Plato \\ 
Aristole}}}
$
}
}
\put(40,-3){
\rm
$\xrightarrow[\text{worldview}]{\text{animistic} }$
$\textcircled{\scriptsize 1}$
}
\put(93,7){
{\line(0,1){25}}
}
\put(93,-7){
{\line(0,-1){35}}
}
}
\put(100,70){
$
\begin{array}{l}
\!\!\!
{\; \xrightarrow[]{ \; 
\quad
}}
\overset{\text{\scriptsize (monism)}}{\underset{\text{\scriptsize (realism)}}
{\fbox{\text{Newton}}}}
{
{\line(1,0){27}}
}
\begin{array}{llll}
\!\!
\rightarrow
{\fbox{\shortstack[l]{theory of \\ relativity}}}
{\xrightarrow[]{\qquad \quad \qquad \qquad \quad}
}
\\
\\
\!\!
\rightarrow
{\fbox{\shortstack[l]{quantum \\ mechanics}}}
{
\xrightarrow[\qquad \qquad \qquad \quad \;\; ]{}
}
\end{array}
\\
\\
\!\! \xrightarrow[]{{ \; 
\quad}}
\overset{\text{\scriptsize (dualism)}}{
\underset{\text{\scriptsize (idealism)}}{\fbox{
{\shortstack[l]{Descartes \\ Kant}}
}}
}
{\xrightarrow[]{
\quad
}}
\underset{\text{\scriptsize (linguistic view)}}{\fbox{
\shortstack[l]{philosophy \\ of language}
}}
\xrightarrow[{
\quad
}
{
\text{
(axiomatization)
}}
]{{
{
\quad
}
{\text{
(quantization)}
}
}}
\end{array}
$
}
\put(330,80){
{\put(-40,0){\drawline(0,-3)(0,-30)}}
{\put(-40,-32){\text{$\xrightarrow[]{\quad  \text{(hint)}}$}}}
{\put(0,-32){\textcircled{\scriptsize 2}}}
}
\put(194,75){\line(0,1){40}}
\put(332,95){
$
\left.\begin{array}{llll}
\; 
\\
\; 
\\
\; 
\\
\;
\end{array}\right\}
\xrightarrow[]{\textcircled{\scriptsize 3}}
\overset{\text{\scriptsize (unsolved)}}{
\underset{\text{\scriptsize (physics)}}{
\fbox{\shortstack[l]{theory of \\ everything}}
}
}
$
}
\put(332,30){
$
\left.\begin{array}{lllll}
\; 
\\
\; 
\\
\; 
\\
\;
\\
\;
\end{array}\right\}
{\xrightarrow[]{\textcircled{\scriptsize 4}}}
\overset{\text{\scriptsize (quantum language)}}{
\underset{\text{\scriptsize (language)}}{
\fbox{\shortstack[l]{measurement \\ theory (=MT)}}
}
}
$
}
\put(2,-20){
{\bf Figure 1. 
\rm
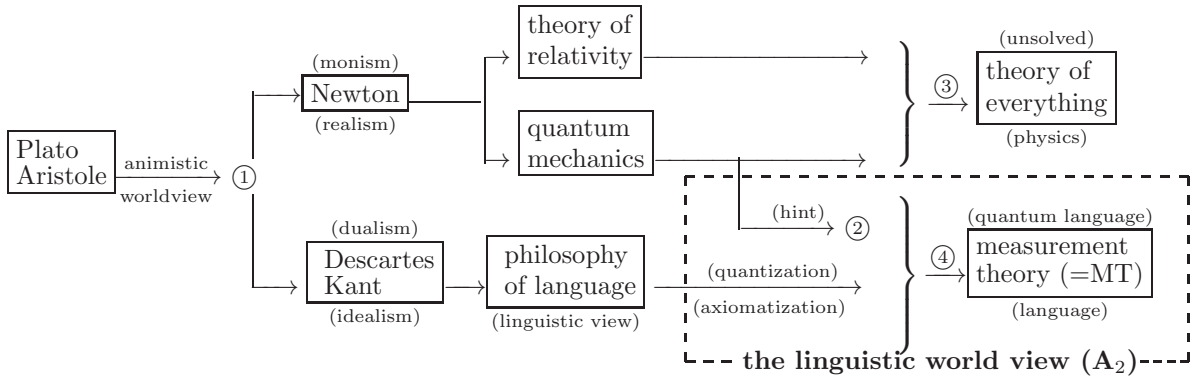
The development of the worldviews.
For the explanations of this figure,
see {{{}}}{\cite{Ishi6,Ishi8}}.
}
}
\put(65,-32){
}
{
\thicklines
\dashline[50]{4}(287,0)(270,0)(270,70)(460,70)(460,0)(442,0)}
{
\put(288,-3){
\bf
the linguistic world view
(A$_2$)
}
}
\end{picture}
\newpage

\par
\noindent
life dwells in thing)
to the mechanical worldview
occurred spontaneously through work of Galileo, bacon, Descartes, Newton
and so on
 (cf. 
\textcircled{\scriptsize 1}
in {\bf Figure 1}).
This power was greatest and opened the door from medieval times 
to modernization. 
Here, the mechanical worldview
means:
\begin{itemize}
\item[(A$_1$)]
Investigate every science (other than physics)
by a model of
mechanics.
\end{itemize}

\newpage
\par
\noindent
Although it is simple, it is now accepted to be the best norm of world description,
and it has reigned over modern science. 
In spite of the great success of the (A$_1$),
in this paper we shall propose
{\lq\lq}the linguistic worldview"
in {\bf Figure 1}.
This says:
\begin{itemize}
\item[(A$_2$)]
Describe every science (other than physics)
by quantum language.
\end{itemize}
(cf. refs. \textcolor{black}{\cite{Ishi1,Ishi2,Ishi3,Ishi4,Ishi5,Ishi6, Ishi7, Ishi8, Ishi9, Ishi10, Ishi11}}).

Note that {\bf Figure 1} does not include the (A$_1$) but the (A$_2$).
Thus, we have an opinion that
the (A$_1$) should not be regarded as the worldview but a hint of scientific thinking.
That is, we think that
the (A$_1$) plays an auxiliary
role in the linguistic world view (A$_2$).

\par
\noindent
\subsection{\normalsize
Quantum language
(=Measurement theory)}
\par

\par
\noindent
It is well known that
quantum mechanics has many interpretations
(i.e.,
the Copenhagen interpretation,
many worlds interpretation, Borm interpretation, probabilistic interpretation,
etc.).
This fact is never desirable.
In 1991,
we proposed the mathematical formulation of Heisenberg's uncertainty principle
(cf.
{\cite{IshiU}}).
However,
we should just have discussed it under a firm interpretation.
Recently 
we proposed the linguistic quantum interpretation
(called quantum and classical measurement theory), which was characterized as a kind of metaphysical and linguistic turn of the Copenhagen interpretation.
This turn from physics to language does not only extend quantum theory to classical systems but also yield the linguistic worldview
(A$_2$)
(i.e., the philosophy of quantum mechanics, in other words,  
quantum philosophy). In fact, we can consider that
traditional philosophies have progressed toward quantum philosophy
(cf. {\bf Figure 1}).
Thus, we expect that the linguistic interpretation
is the only one interpretation of quantum mechanics
( see (G) later).

In this paper, we first review the linguistic quantum interpretation
(in Section 2), 
and further, 
we discuss
inference interval
(in Section 3),
statistical hypothesis test
(in Section 4),
Fisher maximum likelihood method,
Bayes method
and
regression analysis
(in Sections 5 and 6)
in measurement theory.
The essential parts of Sections 5 and 6 were published in
{\cite{Ishi7}}.

\par
The purpose of this paper
is
to answer the question:
\begin{itemize}
\item[(A$_3$)]
What is statistics?
$\;\;$
Or, where is statistics in science?
\end{itemize}
This will be answered in the framework of
{\bf Figure 1}.
\par

\par
\vskip0.3cm
\par
\noindent
\section{\large
Measurement Theory
(Axioms
and
Interpretation)
}
\par
\noindent
\subsection{\normalsize
Mathematical Preparations
}

\rm
\par
\par
\noindent
In this section,
we prepare mathematics,
which is used in measurement theory
(or in short, MT).
\par
\par
\rm
Measurement theory
is,
by a hint of
quantum mechanics
(
i.e.,
the
{\lq\lq}$\xrightarrow[\text{(hint)}]{}$
\textcircled{\scriptsize 2}"
in {\bf Figure 1}), constructed
as the scientific
language
formulated
in a certain 
\textcolor{black}{$C^*$}-algebra ${\cal A}$
(i.e.,
a norm closed subalgebra
in the operator algebra $B(H)$
composed of all bounded linear operators on a Hilbert space $H$,
{\rm cf.$\;$}\textcolor{black}{\cite{Neum, Saka}}
).
MT is composed of
two theories
(i.e.,
pure measurement theory
(or, in short, PMT]
and
statistical measurement theory
(or, in short, SMT).
That is,
we see:
\par
\rm
\par
\begin{itemize}
\item[(B)]
$
\underset{\text{\footnotesize }}{
\text{
MT (measurement theory=quantum language)
}
}
$
\\
$=\cases
\text{(B$_1$)}:
\underset{\text{\scriptsize }}{\text{[PMT
]}}
\\
=
\displaystyle{
{
\mathop{\mbox{[(pure) measurement]}}_{\text{\scriptsize (Axiom$^{\rm P}$ 1) }}
}
}
+
\displaystyle{
\mathop{
\mbox{
[causality]
}
}_{
{
\mbox{
\scriptsize
(Axiom 2)
}
}
}
}
\\
\\
\text{(B$_2$)}
:
\underset{\text{\scriptsize }}{\text{[SMT
]}}
\\
=
\displaystyle{
{
\mathop{\mbox{[(statistical) measurement]}}_{\text
{\scriptsize (Axiom$^{\roman S}$ 1) }}
}
}
\!
+
\!
\displaystyle{
\mathop{
\mbox{
[causality]
}
}_{
{
\mbox{
\scriptsize
(Axiom 2)
}
}
}
}
\endcases
$
\end{itemize}
where
Axiom 2 is common in PMT and SMT.
For completeness, note that measurement theory (B)
(i.e.,
(B$_1$) and (B$_2$))
is
a kind of language
based on
{\lq\lq}the quantum mechanical world view{\rq\rq}.
It may be understandable
to
consider that
\begin{itemize}
\item[(C$_1$)]
PMT and SMT
is related to
Fisher's statistics
and
Bayesian statistics
respectively.
\end{itemize}

%

When ${\cal A}=B_c(H)$,
the ${C^*}$-algebra composed
of all compact operators on a Hilbert space $H$,
the (B) is called {quantum measurement theory}
(or,
quantum system theory),
which can be regarded as
the linguistic aspect of quantum mechanics.
Also, when ${\cal A}$ is commutative
$\big($
that is, 
when ${\cal A}$ is characterized by $C_0(\Omega)$,
the $C^*$-algebra composed of all continuous 
complex-valued functions vanishing at infinity
on a locally compact Hausdorff space $\Omega$
({\rm cf.$\;$}\textcolor{black}{\cite{Saka}})$\big)$,
the (B) is called {classical measurement theory}.
Thus, we have the following classification:
\begin{itemize}
\item[(C$_2$)]
$
\quad
\underset{\text{\scriptsize }}{\text{MT}}
$
$\left\{\begin{array}{ll}
\text{quantum MT$\quad$(when ${\cal A}=B_c (H)$)}
\\
\\
\text{classical MT
$\quad$
(when ${\cal A}=C_0(\Omega)$)}
\end{array}\right.
$
\end{itemize}
%

\par
%

\par
\noindent

\par
\noindent
\par
Now we shall explain the measurement theory
(B).
Let
${\cal A}
( \subseteq B(H))$
be
a
${C^*}$-algebra,
and let
${\cal A}^*$ be the
dual Banach space of
${\cal A}$.
That is,
$ {\cal A }^* $
$ {=}  $
$ \{ \rho \; | \; \rho$
is a continuous linear functional on ${\cal A}$
$\}$,
and
the norm $\| \rho \|_{ {\cal A }^* } $
is defined by
$ \sup \{ | \rho ({}F{}) |  \:{}: \; F \in {\cal A}
\text{ such that }\| F \|_{{\cal A}} 
(=\| F \|_{B(H)} )\le 1 \}$.
The bi-linear functional
$\rho(F)$
is
also denoted by
${}_{{\cal A}^*}
\langle \rho, F \rangle_{\cal A}$,
or in short
$
\langle \rho, F \rangle$.
Define the
\it
mixed state
$\rho \;(\in{\cal A}^*)$
\rm
such that
$\| \rho \|_{{\cal A}^* } =1$
and
$
\rho ({}F) \ge 0
\text{ 
for all }F\in {\cal A}
\text{ satisfying }
F \ge 0$.
And
put
\begin{align*} {\frak S}^m  ({}{\cal A}^*{})
{=}
\{ \rho \in {\cal A}^*  \; | \;
\rho
\text{ is a mixed state}
\}.
\end{align*}
%
\rm
A mixed state
$\rho (\in {\frak S}^m  ({\cal A}^*) $)
is called a
\it
pure state
\rm
if
it satisfies that
{\lq\lq $\rho = \theta \rho_1 + ({}1 - \theta{}) \rho_2$
for some
$ \rho_1 , \rho_2 \in {\frak S}^m  ({\cal A}^*)$
and
$0 < \theta < 1 $\rq\rq}
implies
{\lq\lq $\rho =  \rho_1 = \rho_2$\rq\rq}\!.
Put
\begin{align*} {\frak S}^p  ({}{\cal A}^*{})
{=}
\{ \rho \in {\frak S}^m  ({\cal A}^*)  \; | \;
\rho
\text{ is a pure state}
\},
\end{align*}
which is called a
\it
state space.
\rm
\rm
Riesz's theorem 
({\rm cf.}\textcolor{black}{\cite{Yosi}}) says that
$C_0(\Omega )^*$
$=$
${\cal M}(\Omega )
$
$=$
$\{
\rho \;|\;
\rho
$
is a signed measure on $\Omega$
$
\}$,
${\frak S}^m(C_0(\Omega )^*)$
$=$
${\cal M}_{+1}^m(\Omega )
$
$=$
$\{
\rho \;|\;
\rho
$
is a measure on $\Omega$
such that
$\rho(\Omega)=1$
$
\}$.
Also,
it is well known
({\rm cf.}\textcolor{black}{\cite{Saka}})
that
$ {\frak S}^p  ({}{B_c(H)}^*{})=$
$\{ | u \rangle \! \langle u |
$
(i.e., the Dirac notation)
$
\:\;|\;\:
$
$
\|u \|_H=1 
\}$,
and
$ {\frak S}^p  ({}{C_0(\Omega)}^*{})$
$={\cal M}_{+1}^p(\Omega)$
$=$
$\{ \delta_{\omega_0} \;|\; \delta_{\omega_0}$ is a point measure at
${\omega_0}
\in \Omega
\}$,
where
$ 
\int_\Omega f(\omega) \delta_{\omega_0} (d \omega )$
$=$
$f({\omega_0})$
$
(\forall f
$
$
\in C_0(\Omega))$.
The latter implies that
$ {\frak S}^p  ({}{C_0(\Omega)}^*{})$
can be also identified with
$\Omega$
(called a {\it spectrum}
or
{\it spectrum space})
such as
\begin{align*}
\underset{\text{\scriptsize (state space)}}{{\frak S}^p  ({}{C_0(\Omega)}^*{})}
\ni \delta_{\omega} \leftrightarrow {\omega} \in 
\underset{\text{\scriptsize (spectrum)}}{\Omega}
\tag{1}
\end{align*}

\par
From here,
$C_0(\Omega)$
(or,
commutative unital $C^*$-algebra that includes
$C_0(\Omega)$)
is, for simplicity, denoted by
$C(\Omega)$.
Thus, we put
${\cal A}^*=C(\Omega)^*={\cal M}(\Omega)$,
${\frak S}^m({\cal A}^*)
={\frak S}^m(C(\Omega)^*)={\cal M}_{+1}^m(\Omega)$,
and
${\frak S}^p({\cal A}^*)
={\frak S}^p(C(\Omega)^*)={\cal M}_{+1}^p(\Omega)
\approx
\Omega$.
And,
for any
mixed state
$\nu \in
{\cal M}_{+1}^m(\Omega)$
and
any observable
${\mathsf O}{\; \equiv}(X, {\cal F},$
$F)$ in 
${C(\Omega)}$,
we put:
\begin{align*}
&
\nu(F(\Xi))
=
{}_{{C(\Omega)}^*}
\langle
{\nu{}, F(\Xi)} 
\rangle
{}_{C(\Omega)}
\\
=
&
{}_{{{\cal M}(\Omega)}}
\langle
{\nu{}, F(\Xi)} 
\rangle
{}_{C(\Omega)}
=
\int_\Omega
[F(\Xi)](\omega) \; \nu (d \omega ).
\tag{2}
\end{align*}
Also,
put
$\nu(D)=\int_D \nu(d \omega)$
$(\forall D \in {\cal B}_\Omega$
:
Borel $\sigma$-field
).
In order to avoid the confusion between
$\nu(F(\Xi))
$
in \textcolor{black}{(2)}
and
$\nu(D)$,
we do not use $\nu(F(\Xi))
$.
Also,
for any $\delta_{\omega_0} \in
{\cal M}_{+1}^p(\Omega)
\approx
\Omega$,
we put:
\begin{align*}
&
{}_{{C(\Omega)}^*}
\langle
{\delta_{\omega_0}, F(\Xi)} 
\rangle
{}_{C(\Omega)}
=
{}_{{{\cal M}(\Omega)}}
\langle
{\delta_{\omega_0}, F(\Xi)} 
\rangle
{}_{C(\Omega)}
\\
=
&
\int_\Omega
[F(\Xi)](\omega) \delta_{\omega_0} (d \omega )
=
[F(\Xi)](\omega_0).
\end{align*}

Here, assume
that
the
${C^*}$-algebra
${\cal A}
( \subseteq B(H))$
is unital,
i.e.,
it
has the identity $I$.
This assumption is not unnatural,
since, if $I \notin {\cal A}$,
it suffices to reconstruct the ${\cal A}$ such that it includes 
 ${\cal A}\cup \{I\}$.

According to the noted idea ({\rm cf.}\textcolor{black}{\cite{ Davi}})
in quantum mechanics,
an {\it observable}
${\mathsf O}{\; \equiv}(X, {\cal F},$
$F)$ in 
${{\cal A}}$
is defined as follows:
\par
\par
\begin{itemize}
\item[(D$_1$)]
[Field]
$X$ is a set,
${\cal F}
(\subseteq {\cal P}(X) $,
the power set of $X$)
is a field of $X$,
that is,
{\lq\lq}$\Xi_1, \Xi_2 \in {\cal F}\Rightarrow \Xi_1 \cup \Xi_2 \in {\cal F}${\rq\rq},
{\lq\lq}$\Xi  \in {\cal F}\Rightarrow X \setminus \Xi \in {\cal F}${\rq\rq}.
\item[(D$_2$)]
[Countably additivity]
$F$ is a mapping from ${\cal F}$ to ${{\cal A}}$ 
satisfying:
(a):
for every $\Xi \in {\cal F}$, $F(\Xi)$ is a non-negative element in 
${{\cal A}}$
such that $0 \le F(\Xi) $
$\le I$, 
(b):
$F(\emptyset) = 0$ and 
$F(X) = I$,
where
$0$ and $I$ is the $0$-element and the identity
in ${\cal A}$
respectively.
(c):
for any countable decomposition $\{ \Xi_1,\Xi_2, \ldots \}$
of $\Xi$
$\in {\cal F}$
(i.e., $\Xi_k , \Xi \in {\cal F}$
such that
$\bigcup_{k=1}^\infty \Xi_k = \Xi$,
$\Xi_i \cap \Xi_j= \emptyset
(i \not= j)$),
it holds that
\end{itemize}
\begin{align*}
&
\quad
\lim_{K \to \infty } \rho( F( \bigcup_{k=1}^K \Xi_k ))
=
\rho( F( \Xi ))
\quad
(
\forall \rho \in {\frak S}^m  ({\cal A}^*)
)
\\
&
\quad
\;\;\;
\text{
(i.e., in the sense of weak convergence).
}
\end{align*}
\par
\noindent
\par
\vskip0.3cm
\par
\par
\noindent
{\bf Remark 1.}
By the Hopf extension theorem
({\rm cf.$\;$}\textcolor{black}{\cite{Yosi}}),
we have the mathematical probability space
$(X,$
$ {\overline{\cal F}},$
$ \rho^m (F(\cdot )) \;)$ 
where
${\overline{\cal F}}$
is the smallest $\sigma$-field such that
${{\cal F}} \subseteq {\overline{\cal F}}$.
For the other formulation
(i.e.,
$W^*$-algebraic formulation
),
see
the appendix in \textcolor{black}{\cite{Ishi5}}.

\par

\par
\noindent
\subsection{\normalsize
Pure Measurement Theory
in (B$_1$)
}
\rm
\par
In what follows,
we shall explain PMT
in
(B$_1$).
\par
\rm
With any {\it system} $S$, a $C^*$-algebra 
${\cal A}( \subseteq B(H))$ can be associated in which the 
pure
measurement theory (B$_1$) of that system can be formulated.
A {\it state} of the system $S$ is represented by an element
$\rho (\in {\frak S}^p  ({}{\cal A}^*{}))$
and an {\it observable} is represented by an observable 
${\mathsf{O}}{\; =} (X, {\cal F}, F)$ in ${{\cal A}}$.
Also, the {\it measurement of the observable ${\mathsf{O}}$ for the system 
$S$ with the state $\rho$}
is denoted by 
${\mathsf{M}}_{{{\cal A}}} ({\mathsf{O}}, S_{[\rho]})$
$\big($
or more precisely,
${\mathsf{M}}_{\cal A} ({\mathsf{O}}{\; =} (X, {\cal F}, F), S_{[\rho]})$
$\big)$.
An observer can obtain a measured value $x $
($\in X$) by the measurement 
${\mathsf{M}}_{\cal A} ({\mathsf{O}}, S_{[\rho]})$.
\par
\noindent
\par
The Axiom$^{\rm P}$ 1 presented below is 
a kind of mathematical generalization of Born's probabilistic interpretation of quantum mechanics.
And thus, it is a statement without reality.
\par
\noindent
{\bf{Axiom$^{\rm P}$ 1\;\;
\rm
$[$Pure Measurement$]$}}.
\it
The probability that a measured value $x$
$( \in X)$ obtained by the measurement 
${\mathsf{M}}_{{{\cal A}}} ({\mathsf{O}}$
${ \equiv} (X, {\cal F}, F),$
{}{$ S_{[\rho_0]})$}
belongs to a set 
$\Xi (\in {\cal F})$ is given by
$
\rho_0( F(\Xi) )
$.
\rm

\par
\par
\vskip0.2cm
\par

\par
Next, we explain Axiom 2 in (B).
Let $(T,\le)$ be a tree, i.e., a partial ordered 
set such that {\lq\lq$t_1 \le t_3$ and $t_2 \le t_3$\rq\rq} implies {\lq\lq$t_1 \le t_2$ or $t_2 \le t_1$\rq\rq}\!.
In this paper,
we assume that
$T$ is finite.
Assume that
there exists an element $t_0 \in T$,
called the {\it root} of $T$,
such that
$t_0 \le t$ ($\forall t \in T$) holds.
Put $T^2_\le = \{ (t_1,t_2) \in T^2{}\;|\; t_1 \le t_2 \}$.
The family
$\{ \Phi_{t_1,t_2}{}: $
${\cal A}_{t_2} \to {\cal A}_{t_1} \}_{(t_1,t_2) \in T^2_\le}$
is called a {\it causal relation}
({\it due to the Heisenberg picture}),
\rm
if it satisfies the following conditions {}{(E$_1$) and 
(E$_2$)}.
\begin{itemize}
\item[{\rm (E$_1$)}]
With each
$t \in T$,
a $C^*$-algebra ${\cal A}_t$
is associated.
\item[{\rm (E$_2$)}]
For every $(t_1,t_2) \in T_{\le}^2$, a Markov operator 
$\Phi_{t_1,t_2}{}: {\cal A}_{t_2} \to {\cal A}_{t_1}$ 
is defined
(i.e.,
$\Phi_{t_1,t_2} \ge 0$,
$\Phi_{t_1,t_2}(I_{{\cal A}_{t_2}})$
$
=
$
$
I_{{\cal A}_{t_1}}$
).
And it satisfies that
$\Phi_{t_1,t_2} \Phi_{t_2,t_3} = \Phi_{t_1,t_3}$ 
holds for any $(t_1,t_2)$, $(t_2,t_3) \in T_\le^2$.
\end{itemize}
\noindent
The family of dual operators
$\{ \Phi_{t_1,t_2}^*{}: $
$
{\frak S}^m  ({\cal A}_{t_1}^*)
\to {\frak S}^m  ({\cal A}_{t_2}^*)
\}_{(t_1,t_2) \in T^2_\le}$
is called a
{
\it
dual causal relation}
({\it
due to the Schr\"{o}dinger picture}).
When
$ \Phi_{t_1,t_2}^*{}$
$
(
{\frak S}^p  ({\cal A}_{t_1}^*)
)$
$\subseteq
$
$
(
{\frak S}^p  ({\cal A}_{t_2}^*)
)$
holds for any
$
{(t_1,t_2) \in T^2_\le}$,
the causal relation is said to be
deterministic.

\par
\par
\rm
Now Axiom 2 in the measurement theory (B) is presented
as follows:
\rm
\par
\noindent
{\bf{Axiom 2}
\rm[Causality]}.
\it
The causality is represented by
a causal relation 
$\{ \Phi_{t_1,t_2}{}: $
${\cal A}_{t_2} \to {\cal A}_{t_1} \}_{(t_1,t_2) \in T^2_\le}$.

\rm
\par
\par
\vskip0.2cm
\par
\noindent
\par

\par
\noindent
\par
\noindent
\par
\noindent
\vskip0.2cm
\par

\noindent
\subsection{\normalsize
Linguistic Interpretation
}
Next,
we have to
study how to use the above axioms
as follows.
That is, we present the following interpretation
(F)
[=(F$_1$)--(F$_3$)],
which is characterized as a kind of linguistic turn
of so-called Copenhagen interpretation
({\rm cf.}\textcolor{black}{\cite{Ishi5, Ishi6}}
).
That is,
we propose:
\begin{itemize}
\item[(F$_1$)]
Consider the dualism composed of {\lq\lq}observer{\rq\rq} and {\lq\lq}system( =measuring object){\rq\rq}.
And therefore,
{\lq\lq}observer{\rq\rq} and {\lq\lq}system{\rq\rq}
must be absolutely separated.
\end{itemize}

\par
\noindent
\vskip0.2cm
\par
\rm
\par
\noindent
\vskip0.5cm
\noindent
\unitlength=0.5mm
\begin{picture}(200,82)(15,0)
\put(-8,0)
{
\allinethickness{0.2mm}
\drawline[-40](80,0)(80,62)(30,62)(30,0)
\drawline[-40](130,0)(130,62)(175,62)(175,0)
\allinethickness{0.5mm}
\path(20,0)(175,0)
%
\put(14,-5){
\put(37,50){$\bullet$}
}
\put(50,25){\ellipse{17}{25}}
\put(50,44){\ellipse{10}{13}}
\put(0,44){\put(43,30){\sf \footnotesize{observer}}
\put(42,25){\scriptsize{(I(=mind))}}
}
\put(7,7){\path(46,27)(55,20)(58,20)}
\path(48,13)(47,0)(49,0)(50,13)
\path(51,13)(52,0)(54,0)(53,13)
\put(0,26){
\put(142,48){\sf \footnotesize system}
\put(143,43){\scriptsize (matter)}
}
\path(152,0)(152,20)(165,20)(150,50)(135,20)(148,20)(148,0)
\put(10,0){}
\allinethickness{0.2mm}
\put(0,-5){
\put(130,39){\vector(-1,0){60}}
\put(70,43){\vector(1,0){60}}
\put(92,56){\sf \scriptsize \fbox{observable}}
\put(58,50){\sf \scriptsize }
\put(57,53){\sf \scriptsize \fbox{\shortstack[l]{measured \\ value}}}
\put(80,44){\scriptsize \textcircled{\scriptsize a}interfere}
\put(80,33){\scriptsize \textcircled{\scriptsize b}perceive a reaction}
\put(130,56){\sf \scriptsize \fbox{state}}
}
}
\put(30,-15){\bf
{\bf Figure 2}. 
\rm
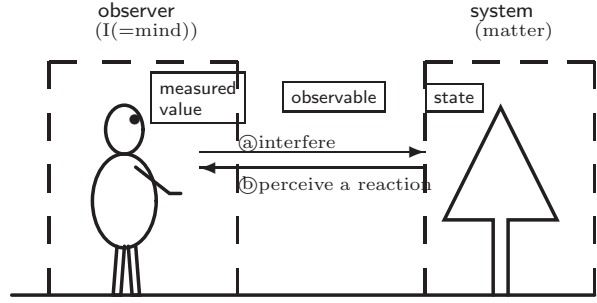
Dualism
in MT
(cf. \cite{Ishi6, Ishi8})
}
\end{picture}
\vskip1.0cm
\par
\noindent

\begin{itemize}
\item[(F$_2$)]
Only one measurement is permitted.
And thus,
the state after a measurement
is meaningless
$\;$
since it 
can not be measured any longer.
Also, the causality should be assumed only in the side of system,
however,
a state never moves.
Thus,
the Heisenberg picture should be adopted,
and thus,
the Schr\"{o}dinger picture
should be prohibited.
\item[(F$_3$)]
Also, the observer
does not have
the space-time.
Thus, 
the question:
{\lq\lq}When and where is a measured value obtained?{\rq\rq}
is out of measurement theory.
And thus,
Schr\"{o}dinger's cat is out of measurement theory,
\end{itemize}
\par
\noindent
and so on.

And therefore, in spite of
Bohr's realistic view,
we propose the following linguistic view:
\begin{itemize}
\item[(G)]
In the beginning was the language called measurement theory
(with the interpretation (F)).
And, for example, quantum mechanics can be
fortunately
described in this language.
And moreover,
almost all scientists have already mastered this language
partially and informally
since statistics
(at least, its basic part)
is characterized as one of aspects of
measurement theory
({\rm cf.} \cite{Ishi3,Ishi4,Ishi7}).
\end{itemize}

\par
\par
\noindent
\noindent
\bf
Remark 2.
\rm
As seen in
the
{\lq\lq}$\xrightarrow[\text{(hint)}]{}$
\textcircled{\scriptsize 2}"
of {\bf Figure 1},
measurement theory was constructed by a hint of quantum mechanics
(or,
by the linguistic turn of quantum mechanics).
However,
in the sense of (G),
it
should be substantially rewritten by
the {\lq\lq}$\xleftarrow[\text{(description)}]{}$
\textcircled{\scriptsize 2}".
Of course, we do not deny that
another quantum physics exists beyond
\textcircled{\scriptsize 3}
in {\bf Figure 1}.

\par
\noindent
\subsection{\normalsize
Sequential Causal Observable
and Its Realization
}

\par
For each
$k=1,$
$2,\ldots,K$,
consider a measurement
${\mathsf{M}}_{{{\cal A}}} ({\mathsf{O}_k}$
${\; \equiv} (X_k, {\cal F}_k, F_k),$
$ S_{[\rho]})$.
However,
since
the (F$_2$)
says that
only one measurement is permitted,
the
measurements
$\{
{\mathsf{M}}_{{{\cal A}}} ({\mathsf{O}_k},S_{[\rho]})
\}_{k=1}^K$
should be reconsidered in what follows.
Under the commutativity condition such that
\begin{align*}
&
F_i(\Xi_i) F_j(\Xi_j) 
=
F_j(\Xi_j) F_i(\Xi_i)
\tag{3}
%
\\
&
\quad
(\forall \Xi_i \in {\cal F}_i,
\forall \Xi_j \in  {\cal F}_j , i \not= j),
\nonumber
\end{align*}
we can
define the product observable
${\text{\large $\times$}}_{k=1}^K {\mathsf{O}_k}$
$=({\text{\large $\times$}}_{k=1}^K X_k ,$
$ \boxtimes_{k=1}^K {\cal F}_k,$
$ 
{\text{\large $\times$}}_{k=1}^K {F}_k)$
in ${\cal A}$ such that
\begin{align*}
({\text{\large $\times$}}_{k=1}^K {F}_k)({\text{\large $\times$}}_{k=1}^K {\Xi}_k )
=
F_1(\Xi_1) F_2(\Xi_2) \cdots F_K(\Xi_K)
\\
\;
(
\forall \Xi_k \in {\cal F}_k,
\forall k=1,\ldots,K
).
\qquad
\qquad
\nonumber
\end{align*}
Here,
$ \boxtimes_{k=1}^K {\cal F}_k$
is the smallest field including
the family
$\{
{\text{\large $\times$}}_{k=1}^K \Xi_k
$
$:$
$\Xi_k \in {\cal F}_k \; k=1,2,\ldots, K \}$.
Then, 
the above
$\{
{\mathsf{M}}_{{{\cal A}}} ({\mathsf{O}_k},S_{[\rho]})
\}_{k=1}^K$
is,
under the commutativity condition \textcolor{black}{(3)},
represented by the simultaneous measurement
${\mathsf{M}}_{{{{\cal A}}}} (
{\text{\large $\times$}}_{k=1}^K {\mathsf{O}_k}$,
$ S_{[\rho]})$.

\par
Consider a tree
$(T{\; \equiv}\{t_0, t_1, \ldots, t_n \},$
$ \le )$
with the root $t_0$.
This is also characterized by
the map
$\pi: T \setminus \{t_0\} \to T$
such that
$\pi( t)= \max \{ s \in T \;|\; s < t \}$.
Let
$\{ \Phi_{t, t'} : {\cal A}_{t'}  \to {\cal A}_{t}  \}_{ (t,t')\in
T_\le^2}$
be a causal relation,
which is also represented by
$\{ \Phi_{\pi(t), t} : {\cal A}_{t}  \to {\cal A}_{\pi(t)}  \}_{ 
t \in T \setminus \{t_0\}}$.
Let an observable
${\mathsf O}_t{\; \equiv}
(X_t, {\cal F}_{t}, F_t)$ in the ${\cal A}_t$ 
be given for each $t \in T$.
Note that
$\Phi_{\pi(t), t}
{\mathsf O}_t$
$(
{\; \equiv}
(X_t, {\cal F}_{t},
\Phi_{\pi(t), t} F_t)$
)
is an observable in the ${\cal A}_{\pi(t)}$.

The pair
$[{\mathbb O}_T]
$
$=$
$[
\{{\mathsf O}_t \}_{t \in T}$,
$\{ \Phi_{t, t'} : {\cal A}_{t'}  \to {\cal A}_{t}  \}_{ (t,t')\in
T_\le^2}$
$]$
is called a
{\it sequential causal observable}.
For each $s \in T$,
put $T_s =\{ t \in T \;|\; t \ge s\}$.
And define the observable
${\widehat{\mathsf O}}_s
\equiv ({\text{\large $\times$}}_{t \in T_s}X_t, \boxtimes_{t \in T_s}{\cal F}_t, {\widehat{F}}_s)$
in ${\cal A}_s$
as follows:
\par
\noindent
\begin{align}
\widehat{\mathsf O}_s
&=
\left\{\begin{array}{ll}
{\mathsf O}_s
\quad
&
\!\!\!\!\!\!\!\!\!\!\!\!\!\!\!\!\!\!
\text{(if $s \in T \setminus \pi (T) \;${})}
\\
{\mathsf O}_s
{\text{\large $\times$}}
({}\bigtimes_{t \in \pi^{-1} ({}\{ s \}{})} \Phi_{ \pi(t), t} \widehat {\mathsf O}_t{})
\quad
&
\!\!\!\!\!\!
\text{(if $ s \in \pi (T) ${})}
\end{array}\right.
\nonumber
\\
&
\!\!\!\!\!
\tag{4}
\end{align}
if
the commutativity condition holds
(i.e.,
if the product observable
${\mathsf O}_s
{\text{\large $\times$}}
({}\bigtimes_{t \in \pi^{-1} ({}\{ s \}{})} \Phi_{ \pi(t), t}
$
$\widehat {\mathsf O}_t{})$
exists)
for each $s \in \pi(T)$.
Using \textcolor{black}{(4)} iteratively,
we can finally obtain the observable
$\widehat{\mathsf O}_{t_0}$
in ${\cal A}_{t_0}$.
The
$\widehat{\mathsf O}_{t_0}$
is called the realization
(or,
realized causal observable)
of
$[{\mathbb O}_T]$.

\par

\noindent

\subsection{\normalsize
Statistical Measurement Theory
in (B$_2$)
}

\rm


We shall introduce the following notation:
\rm
It is usual to consider that
we do not know the pure state
$\rho_0^p$
$(
\in
{\frak S}^p  ({}{\cal A}^*{})
)$
when
we take a measurement
${\mathsf{M}}_{{{\cal A}}} ({\mathsf{O}}, S_{[\rho_0^p]})$.
That is because
we usually take a measurement ${\mathsf{M}}_{{{\cal A}}} ({\mathsf{O}},
S_{[\rho_0^p]})$
in order to know the state $\rho_0^p$.
Thus,
when we want to emphasize that
we do not know the state $\rho_0^p$,
${\mathsf{M}}_{{{\cal A}}} ({\mathsf{O}}, S_{[\rho_0^p]})$
is denoted by
${\mathsf{M}}_{{{\cal A}}} ({\mathsf{O}}, S_{[\ast]})$.
Also,
when we know the distribution $\rho_0^m$
$( \in {\frak S}^m({\cal A}^*) )$
of the unknown state
$\rho_0^p$,
the
${\mathsf{M}}_{{{\cal A}}} ({\mathsf{O}}, S_{[\rho_0^p]})$
is denoted by
${\mathsf{M}}_{{{\cal A}}} ({\mathsf{O}}, S_{[\ast]}
(\{ \rho_0^m \}) )$.
The $\rho_0^m$
is called a mixed state.
And further,
if
we know that a mixed state $\rho_0^m$
belongs to
a compact set $K$
$(\subseteq 
{\frak S}^m({\cal A}^*) )$,
the
${\mathsf{M}}_{{{\cal A}}} ({\mathsf{O}}, S_{[\rho_0^p]})$
is denoted by
${\mathsf{M}}_{{{\cal A}}} ({\mathsf{O}}, S_{[\ast]}
(K) )$.

\par
\vskip0.3cm

\par
The Axiom$^{\rm S}$ 1 presented below is 
a kind of mathematical generalization of 
Axiom$^{\rm P}$ 1.

\par

\par
\noindent
{\bf{Axiom$^{\rm S}$\;1\;
\rm
\;[Statistical measurement]}}.
\it
The probability that a measured value $x$
$( \in X)$ obtained by the measurement 
${\mathsf{M}}_{{{\cal A}}} ({\mathsf{O}}$
${ \equiv} (X, {\cal F}, F),$
{}{$ S_{[\ast]}(\{ \rho_0^m \}) )$}
%
belongs to a set 
$\Xi (\in {\cal F})$ is given by
$
\rho_0^m ( F(\Xi) )
$
$($
$=
{}_{{{\cal A}^*}}\langle
\rho_0^m,
F(\Xi)
\rangle_{{\cal A}}$
$)$.
\rm


\par

Thus, we can propose the statistical measurement theory
(B$_2$),
in which
Axiom 2 and Interpretation 
(G) are common.

Let
$\widehat{\mathsf O} $
$\equiv$
$(X \times Y, {\cal F} \bigstimes {\cal G}, {H})$
be an observable
in
a
$C^*$-algebra
${\cal A}$.
Assume that
we know that the measured value $(x,y) \;(\in X \times Y)$
obtained by a statistical measurement
${\mathsf M}_{\cal A}(\widehat{\mathsf O}, S_{[*]}(\{
\rho_0^m
\}))$ belongs to
$\Xi \times Y \;(\in {\cal F} \bigstimes {\cal G})$.
Then,
there is a reason to infer that the unknown measured value
$y \;(\in Y)$ is distributed under the conditional probability
$
P_\Xi (G(\Gamma))
$, where
\begin{align*}
P_\Xi (G(\Gamma))
&
= \frac{ 
{{}_{{\cal A}^*}}\langle
{\rho_0^m, {H}
(\Xi \times \Gamma)} 
\rangle_{\cal A}
}
{ 
{{}_{{\cal A}^*}}\langle {\rho_0^m, {H}(\Xi \times Y)} 
\rangle_{\cal A}}
\quad (\forall \Gamma \in {\cal G})
\tag{5} 
\end{align*}

\par

\par
\par
Thus,
by
a hint of
Fisher's maximum likelihood method,
we have the following theorem,
which is the most fundamental
in this paper.
\par
\noindent
{\bf Theorem 1}
\rm
[{}Fisher's maximum likelihood method
in general ${\cal A}$
(cf. \cite{Ishi7}].
\it
Let
$\widehat{\mathsf O} $
$\equiv$
$(X \times Y, {\cal F} \bigstimes {\cal G}, {H})$
be an observable
in
a
$C^*$-algebra
${\cal A}$.
Let
$K
(\subseteq {\frak S}^m({\cal A}^*))$
be a compact set.
Assume that
we know that the measured value $(x,y) \;(\in X \times Y)$
obtained by a measurement
${\mathsf M}_{\cal A}(\widehat{\mathsf O}, S_{[*]}(K))$ belongs to
$\Xi \times Y \;(\in {\cal F} \bigstimes {\cal G})$.
Then,
there is a reason to infer that the unknown measured value
$y \;(\in Y)$ is distributed under the conditional probability
$
P_\Xi (G(\Gamma))
$, where
\begin{align*}
P_\Xi (G(\Gamma))
&
= \frac{ 
{}_{{\cal A}^*}
\langle
{\rho_0^m, {H}
(\Xi \times \Gamma)
\rangle_{\cal A} 
}
}
{
{}_{{\cal A}^*}
\langle
{\rho_0^m, {H}
(\Xi \times Y)
\rangle_{\cal A} 
}
}
\quad (\forall \Gamma \in {\cal G}).
\tag{6} 
\end{align*}
Here, $\rho_0^m \;(\in K \subseteq {\frak S}^m ({\cal A}^*))$
is defined by
\begin{align*}
{
{}_{{\cal A}^*}
\langle {\rho_0^m, {H}(\Xi\times Y)\rangle_{\cal A}
}
}
= \max_{\rho^m \in K 
}
{{}_{{\cal A}^*}} \langle {\rho^m, {H}(\Xi\times Y)
\rangle_{\cal A}
}.
\end{align*}

\par 

\rm
\par
\vskip0.3cm
\par

\noindent
{\bf Corollary 1}.
\it
Let
$(X , {\cal F} , F )$
and
$(Y, {\cal G} , G )$
be observables
in
a
$C^*$-algebra
${\cal A} $.
Let
$\widehat{\mathsf O} $
$\equiv$
$(X \times Y, {\cal F} \bigstimes {\cal G}, F \otimes G)$
be the tensor observable
in
a
tensor
$C^*$-algebra
${\cal A} \otimes {\cal A}$.
Let
$K
(\subseteq {\frak S}^p({\cal A}^*))$
be a compact set.
And put
$K^2_D = \{ \rho \otimes \rho \;| \; \rho
\in K \}$.
Assume that
we know that the measured value $(x,y) \;(\in X \times Y)$
obtained by a measurement
${\mathsf M}_{{\cal A}\otimes {\cal A}}
(\widehat{\mathsf O}, S_{[*]}(K_D^2))$ belongs to
$\Xi \times Y \;(\in {\cal F} \bigstimes {\cal G})$.
Then,
there is a reason to infer that the unknown measured value
$y \;(\in Y)$ is distributed under the conditional probability
$
P_\Xi (G(\Gamma))
$, where
\begin{align*}
P_\Xi (G(\Gamma))
&
= 
{}_{{\cal A}^*}
\langle
{\rho_0^p, {G}
( \Gamma)
\rangle_{\cal A} 
}
\quad (\forall \Gamma \in {\cal G}).
\tag{7} 
\end{align*}
Here, $\rho_0^p \;(\in K \subseteq {\frak S}^p ({\cal A}^*))$
is defined by
\begin{align*}
{
{}_{{\cal A}^*}
\langle {\rho_0^p, {F}(\Xi)\rangle_{\cal A}
}
}
= \max_{\rho^p \in K 
}
{{}_{{\cal A}^*}} \langle {\rho^p, {F}(\Xi)
\rangle_{\cal A}
}.
\end{align*}
\par
\noindent
{\it Proof.}
\rm
The (7) is,
from (6), 
derived as follows:
\begin{align*}
P_\Xi (G(\Gamma))
&
= \frac{ 
{}_{{\cal A}^* \otimes {\cal A}^*}
\langle
{\rho_0^p \otimes \rho^p_0, {(F \otimes G)}
(\Xi \times \Gamma)
\rangle_{\cal A \otimes {\cal A}} 
}
}
{
{}_{{\cal A}^* \otimes {\cal A}^*}
\langle
{\rho_0^p \otimes \rho^p_0, {(F \otimes G)}
(\Xi \times Y)
\rangle_{\cal A \otimes {\cal A}} 
}
}
\\
&
=
{}_{{\cal A}^*}
\langle
{\rho_0^p, {G}
( \Gamma)
\rangle_{\cal A} 
}
\end{align*}
\qed

\par 

\rm
\par
\vskip0.3cm
\par
\par
\noindent
{\bf Remark 3}
[The state before a measurement].
In the sense of Corollary 1,
the arbitrariness of
$(Y, {\cal G} , G )$
says that,
when we  know that the measured value obtained by
${\mathsf M}_{\cal A}( {\mathsf O} =(X, {\cal F}, F), S_{[\ast]})$
belongs to
$\Xi ( \in {\cal F})$,
we can infer that
the state $[\ast]$ (before the measurement)
is $\rho_0^p$.

\rm

\noindent
\par


\vskip1.0cm
\par
\par
\rm
\section{Inference interval }
\par
\noindent
\par
Let
${\mathsf O} ({}\equiv ({}X, {\cal F} , F{}){})$
be an observable
%
formulated in a
$C^*$-algebra
${\cal A}$.
Assume that
$X$ has a metric $d_X$.
And assume
that
the state space
${\frak S}^p ({\cal A}^*)$
has the metric $d_{\frak S}$,
which induces the
weak$^*$ topology
$\sigma({}{\cal A}^* , {\cal A}{})$.
\index{estimator@estimator}
Let
$E:X \to {\frak S}^p ({\cal A}^*)$
be a continuous map,
which is called
\it
{\lq\lq} estimator{\rq\rq}$\!\!\!\!.\; \;$
\rm
Let
$\gamma$
be a real number such that
$0 \ll \gamma < 1$,
for example,
$\gamma = 0.95$.
For any
$ \rho^{p}({}\in{\frak S}^p ({\cal A}^*))$,
define
the positive number
$\eta^\gamma_{\rho^{p}}$
$({}> 0)$
such that:
\begin{align*}
\eta^\gamma_{\rho^{p}}
=
\inf
\{
\eta > 0:
{}_{{}_{{\cal A}^* } }
\Bigl\langle \rho^p , F ({}E^{-1} ({}
B(\rho^p ; \eta{}))  \Bigl\rangle{}_{{}_{{\cal A} }}
\ge \gamma
\}
\end{align*}
where
$B(\rho^p ; \eta)$
$=$
$\{ \rho_1^p
({}\in{\frak S}^p ({\cal A}^*)):
d_{\frak S} ({}\rho_1^p, \rho^p{}) \le \eta \}$.
For any
$x$
$({}\in X{})$,
put
\begin{align*}
D_x^{\gamma}
=
\{
{\rho^{p}}
(\in
{\frak S}^p ({\cal A}^*))
:
d_{\frak S} ({}E(x),
\rho^{p})
\le
\eta^\gamma_{\rho^p }
\}.
\tag{8} \end{align*}
The $D_x^{\gamma}$ is called
\it
the $({}\gamma{})$-inference interval
of the
measured value
$x$.
\rm
\par
Note that,
\begin{enumerate}
\item[(H)]
\it
for any
$\rho_0^{p}
({}\in
{\frak S}^p ({\cal A}^*))$,
the probability,
that
the measured value $x$
obtained
by the measurement
${\mathsf M}_{\cal A} \big({}{\mathsf O}:= ({}X, {\cal F} , F{})  ,$
$ S_{[\rho_0^{p} {}] } \big)$
satisfies the following
condition $(\flat)$,
is larger than
$\gamma$
({}e.g., $\gamma= 0.95${}).
\begin{enumerate}
\item[$(\flat)$]
$ E(x) \in B({}\rho_0^{p} ; {\eta }_{\rho_0^p}^\gamma{}) $
or equivalently
\\
$  d(E(x),  \rho_0^{p}{}) \le  {\eta }^\gamma_{\rho_0^p}  $.
\end{enumerate}
\end{enumerate}
\par
\noindent
Assume that
we get
a measured value
$x_0$
by
the measurement
${\mathsf M}_{\cal A} \big({}{\mathsf O}:= ({}X, {\cal F} , F{})  ,$
$ S_{[\rho_0^{p} {}] } \big)$.
Then,
we see the following equivalences:
$$
(\flat) \; \Longleftrightarrow \;
d_{\frak S}({}E({}x_0{}), \rho_0^p{}) \le \eta^\gamma_{\rho_0^p }
\;
\Longleftrightarrow
\;
D_{x_0}^\gamma
\ni
\rho_0^p.
$$

\par
\noindent
\unitlength=0.4mm
\begin{picture}(100,75)
\put(-9,0){{
\put(40,16){\scriptsize $x_0$}
\qbezier(40,20)(100,61)(157,42)
\path(107,49)(115,48)(107,45)
\put(40,20){\circle*{1}}
\put(157,41){\circle*{1}}
\put(155,45){\scriptsize $E({}x_0)$}
\put(151,33){$ \; \omega_0$}
\put(149,34){ \circle*{1} }
\put(170,30){$ D_{x_0}^\gamma$}
\put(153,63){ $\Omega$}
\put(57,63){$X$}
\allinethickness{0.5mm}
\put(60,30){\oval(70,60)}
\put(160,30){\oval(70,60)}
\allinethickness{0.3mm}
\put(157,42){\ellipse{30}{30}}
}}
\put(40,-20){\bf Figure 3.
\rm
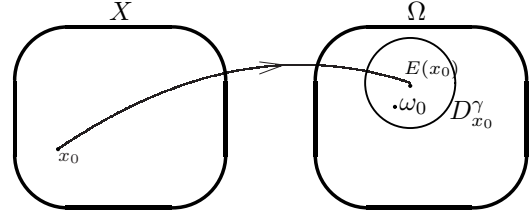
Inference interval
$D^\gamma_{x_0}$
}
\end{picture}

\par
\vskip1.0cm
\par

\par
Summing the above argument,
we have the following theorem.
\par
\noindent
\bf 
Proposition 1
\rm
[{}Inference interval{}].
\index{Inference interval@inference interval}
\sl
Let
${\mathsf O}:= ({}X, {\cal F} , F{}) $
be an observable in
${\cal A}$.
Let
$\rho_0^{p}$
be any fixed state,
i.e.,
$\rho_0^{p} \in
{\frak S}^p ({\cal A}^*)
$,
Consider
a measurement
${\mathsf M}_{\cal A} \big({}{\mathsf O}:= ({}X, {\cal F} , F{})  ,$
$ S_{[\rho_0^{p} {}] } \big)$.
Let
$E:X \to {\frak S}^p ({\cal A}^*)$
be an
estimator.
Let
$\gamma$ be such as $0 \ll \gamma < 1$
({}e.g., $\gamma = 0.95${}).
For any
$x
({}\in X{})$,
define
$D_x^{\gamma}$
as in (8).
Then, we see,
\begin{enumerate}
\item[$(\sharp)$]
the probability
that
the measured value
$x_0
({}\in X)$
obtained
by the measurement
${\mathsf M}_{\cal A} \big({}{\mathsf O}:= ({}X, {\cal F} , F{})  ,$
$ S_{[\rho_0^{p} {}] } \big)$
satisfies
the condition
that
\begin{align*}
\text{
$D_{x_0}^{\gamma} \ni \rho_0^p $
},
\end{align*}
is larger than
$\gamma$.
\end{enumerate}
\par

\rm
\par
\noindent
\bf 
Example 1
\rm
\rm
[{}The urn problem{}].
Put
$\Omega$
$=$
$[{}0, 1{}]$,
i.e.,
the closed interval in ${\mathbb R}$.
We assume that
each
$\omega$
$({}\in \Omega \equiv [{}0, 1{}]{})$
represents an urn that contains a lot of
red balls
and white balls
such that:
\begin{align*}
&
\frac{
\text{ the number of white balls in the urn $\omega$}
}
{
\text{ the total number of balls in the urn $\omega$}
}
\\
\approx
&
\;\;
\omega
\quad
({}\forall \omega \in [0,1] \equiv \Omega{}).
\end{align*}
Define the observable
${\mathsf O} = ({} X \equiv \{ r, w \}, {\cal P}({\{ r, w \}  })  , F{})$
in
$C({}\Omega{})$
such that
where
\begin{align*}
&
F({}\emptyset{})(\omega)= 0, \quad
F({}\{ r \}{})(\omega)   = \omega, \quad 
\\
&
F({}\{ w \}{})(\omega{})
= 1- \omega ,\quad
F({}\{ r, w \}{})(\omega)= 1
\\
& \qquad
({}\forall \omega \in [{}0, 1{}] \equiv \Omega{}).
\end{align*}
\par
\noindent
Here,
consider the following measurement
$M_\omega$:
\begin{align*}
M_\omega
&
:=
\text
{
{\lq\lq} Pick out one ball from the
urn $\omega$,}
\\
&
\text{
and recognize the color of the ball{\rq\rq}}
\end{align*}
That is, we consider
\begin{align*}
M_\omega =
{\mathsf M}_{C ({}\Omega{}) } ({}{\mathsf O} ,
S_{ [{}\delta_{\omega}]}{}).
\end{align*}
Moreover we define the product observable
${\mathsf O}^N$
$\equiv$
$({}X^N , {\cal P}({}X^N{}) , F^N{})$,
such that:
\begin{align*}
&
[
F^N ({}\Xi_1 \bigtimes \Xi_2 \bigtimes \cdots \bigtimes \Xi_{N-1} \bigtimes \Xi_{N}{})
]
({}\omega{})
\\
=
&
[F({}\Xi_1{})]({}\omega{}) \cdot
[F({}\Xi_2{})]({}\omega{}) \cdots
[F({}\Xi_N{})]({}\omega{})
\\
&
(\forall
\omega
\in \Omega \equiv [0,1],
\quad
\forall
\Xi_1, \Xi_2, \cdots , \Xi_N \subseteq X \equiv \{ r,  w \}).
\end{align*}
Note that
\begin{align*}
&
\text{
{\lq\lq} take a measurement $M_\omega$ N times{\rq\rq}}
\\
\Leftrightarrow
&
\text{
{\lq\lq} take a measurement
${\mathsf M}_{C(\Omega{})} ({}{\mathsf O}^{N} , S_{[\delta_{\omega}]}{})${\rq\rq}
}
\end{align*}
Define the estimator
$E: X^{N} ({}\equiv \{ r,w \}^{N}{}) \to \Omega ({}\equiv [0,1]{})$
\par
\noindent
\begin{align*}
&
E({}x_1, x_2 , \cdots , x_{N-1} , x_{N}{})
\\
=
&
\frac{
\sharp [{}
\{ n \in \{ 1,2, \cdots , N \} \; | \; x_n = r \}{}]
}
{N}
\\
&
\quad
(\forall
x=
({}x_1, x_2 , \cdots , x_{N-1} , x_{N} {})
\in X^{N}
\equiv
\{r,w \}^{N}).
\tag{9} \end{align*}
For each
$\omega ({}\in [0,1] \equiv \Omega{})$,
define the positive number
$\eta^\gamma_\omega$
such that:
\begin{align*}
&
\eta^\gamma_\omega
\\
=
&
\inf
\Big\{
\eta > 0 \; \Big| \;
[
F^{N}
(\{ ({}x_1, x_2 , \cdots , x_{N}{})
\; | \;
\omega - \eta
\\
\le
&
E({}x_1, x_2 , \cdots , x_{N}{})
\le
\omega+ \eta
\})]
({}\omega{})
>
0.95
\Big\}
\\
=
&
\displaystyle{
\mathop
{\text{\Large inf}}_{
[
F^{N}
(\{ ({}x_1, x_2 , \cdots , x_{N}{})
:
|
E({}x_1, x_2 , \cdots , x_{N}{})
-
\omega
|
\le
\eta
\})
]
({}\omega{})
>
0.95
}
}
\eta.
\end{align*}
Put
\begin{align*}
D_x^\gamma
=
\{
\omega
({}\in \Omega{})
:
\;
| E({}x) - \omega | \le \eta_\omega^\gamma \}.
\end{align*}
For example,
assume that
$N$ is sufficiently large and
$\gamma = 0.95$.
Then we see, from the property of binomial distribution, that
\begin{align*}
\eta_\omega^{0.95}
\approx
1.96
\sqrt{ \frac{\omega ({}1- \omega{})}{N} }
\intertext{and}
D_x^{0.95}
=
[{}E(x) -\eta_- , E(x) + \eta_+{}]
\end{align*}
where
\begin{align*}
\eta_-
=
\eta_{ E(x) - \eta_- }^{0.95},
\quad
\eta_+
=
\eta_{ E(x) + \eta_+ }^{0.95}.
\end{align*}
Under the assumption that
$N$ is sufficiently large,
we can consider that
$$
\eta_-
\approx
\eta_+
\approx
\eta_{E(x)}^{0.95}
\approx
1.96
\sqrt{ \frac{E(x) ({}1- E(x){})}{N} }.
$$
Then we can conclude that
\begin{itemize}
\item[(I)]
\it
for any
urn
$\omega
({}\in
\Omega \equiv [0,1]))$,
the probability,
that
the measured value $x=$
$(x_1, x_2 , \cdots ,$
$ x_{N}{})$
obtained
by the measurement
${\mathsf M}_{\cal A} \big({}{\mathsf O}^{N}   ,$
$ S_{[\delta_{\omega }{}] } \big)$
satisfies the following
condition $(\sharp{})$,
is larger than
$\gamma$
({}e.g., $\gamma= 0.95${}).
\begin{itemize}
\item[$(\sharp{})$]
$ | \omega - E(x) |
\le
1.96
\sqrt{ \frac{E(x) ({}1- E(x){})}{N} } 
\le
\frac{0.98}{\sqrt{N}}
$.
\end{itemize}
\end{itemize}


\par
\par
\noindent
\section{
\large
Statistical Hypothesis Testing
}

\par
\par
\noindent
\par
\rm

\noindent
\subsection{\normalsize
Problem
(Statistical Hypothesis Testing)
}

\rm
\rm
\par
\noindent
\par
It is usual to consider that
we do not know the pure state
$\rho_0$
$(
\in
{\frak S}^p  ({}{\cal A}^*{})
)$
when
we take a measurement
${\mathsf{M}}_{{{\cal A}}} ({\mathsf{O}}, S_{[\rho_0]})$.
That is because
we usually take a measurement ${\mathsf{M}}_{{{\cal A}}} ({\mathsf{O}},
S_{[\rho_0]})$
in order to know the state $\rho_0$.
Thus,
when we want to emphasize that
we do not know the state $\rho_0$,
${\mathsf{M}}_{{{\cal A}}} ({\mathsf{O}}, S_{[\rho_0]})$
is denoted by
${\mathsf{M}}_{{{\cal A}}} ({\mathsf{O}}, S_{[\ast]})$.

\rm
\par
\noindent
\par
\vskip0.5cm
\par
\noindent
\par
In what follows
we shall study
{\lq\lq}statistical hypothesis testing{\rq\rq}$\!\!\!.\;$
\rm
Consider a measurement
%
${\mathsf M}_{\cal A}({\mathsf O}\equiv(X, {\cal F}, F{}), S_{[*]}
)$
formulated in
${\cal A}$.

Here, we assume that
$(X,
\tau{{}_X})$ is a topological space,
where
$\tau{{}_X}$
is the set of all open sets.
And assume that
$\overline{\cal F}={\cal B}_X$;
the Borel field,
i,e.,
the smallest $\sigma$-field that contains all
open sets in $X$.
Note that we can assume, without loss of generality,
that
$
F({\Xi})
\not=
0
$
for any open set $\Xi (\in \tau{{}_X} )$
such that $\Xi \not= \emptyset$.
That is because,
if
$
F({\Xi})
=
0
$,
it suffices to
redefine
$X$ by $X\setminus \Xi$.

Assume the following hypothesis called {\lq\lq}{\it null hypothesis}":
\begin{itemize}
\item[\textcolor{black}{(J)}]
the unknown state
$[\ast]$
belongs to
a set ${\mathcal N}_H$
$({}\subseteq 
{\frak S}^p({\cal A}^*)
)$.
\end{itemize}

%

\par
\noindent
In order to deny this hypothesis \textcolor{black}{(J)},
we define
the rejection region $
{\widehat R}^\alpha_{{\mathcal N}_H}
$
($\in \overline{\cal F}$) as follows.

\begin{itemize}
\item[\textcolor{black}{\roman{(K)}}]
For sufficiently small
{\it significance level}
$\alpha$
(
$0 < \alpha \ll 1$
,
e.g.,
$\alpha=0.05$
),
define the
{\it
rejection region
}
${\widehat R}^\alpha_{{\mathcal N}_H} \in \overline{\cal F}$
such that
\begin{itemize}
\item[(K$_1$)]
$
{{}_{{\cal A}^*}} \langle 
\rho, 
F({\widehat R}^\alpha_{{\mathcal N}_H})
\rangle_{{}_{\cal A}}
\le
\alpha
\quad
(\forall 
\rho \in {\mathcal N}_H \subseteq {\frak S}^p({\cal A}^* )
)
$
\item[(K$_2$)]
If 
${\widehat R}^{\alpha,1}_{{\mathcal N}_H}
(\in \overline{\cal F} )
$
and
${\widehat R}^{\alpha,2}_{{\mathcal N}_H}
(\in \overline{\cal F} )$
satisfy
(K$_1$)
and
${\widehat R}^{\alpha,1}_{{\mathcal N}_H} \subsetneq 
{\widehat R}^{\alpha,2}_{{\mathcal N}_H}
$,
then,
choose
${\widehat R}^{\alpha,2}_{{\mathcal N}_H}
$.
\end{itemize}
\end{itemize}

\par
\noindent
%

\unitlength=0.20mm
\begin{picture}(400,130)
\put(-30,0){{
\put(27,18){0}
\put(27,108){1}
\put(350,18){$
{\frak S}^p({\cal A}^* )
$}
\dottedline{3}(40,110)(340,110)
\put(40,20){\line(0,1){100}}
\thicklines
\put(40,20){\line(1,0){300}}
\thicklines
\spline(40,110)(60,109)(80,105)(110,100)
(150,30)(180,23)(200,27)(250,25)
(270,35)(280,70)(300,100)(340,110)
\dottedline{5}(40,40)(340,40)
\put(27,40){$\alpha$}
\multiput(150,18)(3,0){40}{\line(0,2){4}}
\put(200,2){${{\mathcal N}_H} $}
\put(134,80){$
{{}_{{\cal A}^*}} \langle 
\rho, 
F(
{\widehat R}^\alpha_{{\mathcal N}_H}
)
\rangle_{{}_{\cal A}}
$}
}}
\put(40,-20){\bf Figure 4.
\rm
Null Hypothesis 
${{\mathcal N}_H}$
}
\end{picture}
\vskip0.5cm
Then, Axiom$^{\rm P}\;$1 says that
\begin{itemize}
\item[(L)]
if
$[\ast] \in {\mathcal N}_H$,
the probability that
a measured value
obtained by
${\mathsf M}_{{\cal A}}({\mathsf O} $
$\equiv(X, {\cal F}, F{}) , S_{[\ast]})$
belong to
${\widehat R}^\alpha_{{\mathcal N}_H}$
is less that
$\alpha$.
Therefore,
if a measured value belongs to
${\widehat R}^\alpha_{{\mathcal N}_H}$,
and
if $\alpha$ is sufficiently small,
then
there is a reason to deny the hypothesis
\textcolor{black}{(J)}.
\end{itemize}


\par
\noindent
It is clear
that
the rejection region ${\widehat R}^\alpha_{{\mathcal N}_H}$
is not uniquely determined
in general.
%
%
Thus, we have the following problem:
\begin{itemize}
\item[(M$_1$)]
Find the most proper rejection region
${\widehat R}^\alpha_{{\mathcal N}_H}$.
\end{itemize}
This will be answered in the following section.

\par
\noindent
\bf
Remark 4.
\rm
Define the observable
${\mathsf O}_1=
(\{0, 1\}, {\cal P}({\{0, 1\}}), G)$
in ${\cal A}$
such that
$\rho (G (\{ 1\}))= 0$
$(\forall \rho \in {\frak S}^p({\cal A}^*) \setminus {\mathcal N}_H)$.
Consider the measurement
${\mathsf M}_{{\cal A} \otimes {\cal A}}
({\mathsf O}_1 \otimes {\mathsf O}, S_{[\ast \otimes \ast ]})$,
where
$\ast \otimes \ast \in \{ \rho \otimes \rho \; |\; \rho \in 
{\frak S}^p ({\cal A}^*) \}$.
Then we see that
the measured value obtained by
${\mathsf M}_{{\cal A} \otimes {\cal A}}
({\mathsf O}_1 \otimes {\mathsf O}, S_{[\ast \otimes \ast ]})$
belongs to
$\{ 0\}  \times {\widehat R}^\alpha_{{\mathcal N}_H}$
is less that $\alpha$.
It is interesting to see the similarity between
statistical hypothesis testing
and
fuzzy contraposition
(cf. \cite{Ishi1}).

\noindent
\subsection{\normalsize
Answer;
Likelihood Ratio Test
}

\rm


\par
\par
\noindent
\par

\par
\noindent
\rm
\par
Let
${\mathsf O} $
$\equiv$
$(X , {\cal F} , {F})$
be an observable
in
a
$C^*$-algebra
${\cal A}$.
Define
the map
$\widehat{L}: \tau{{}_X} \times 
{\frak S}^p({\cal A}^*)
\to
[0,1]$
such that
$$
\widehat{L}
(\Xi
, \rho )
=
\frac{ 
{}_{{\cal A}^*}
\langle
{\rho, {F}
(\Xi )
\rangle_{\cal A} 
}
}
{
\underset{\rho \in {\frak S}^p({\cal A}^*) }{\roman \large sup}
{}_{{\cal A}^*}
\langle
{\rho, {F}
(\Xi )
\rangle_{\cal A} 
}
}.
$$
Further,
define the likelihood function
$L: X \times {\frak S}^p({\cal A}^*)$
$\to [0, \; 1]$
such that
$$
L( x, \rho )
=
\lim_{\overline{\cal F} \ni \Xi \to x }
\widehat{L}(\Xi
, \rho ).
$$
That is, for any positive $\epsilon$,
there exists
an open set
$\Xi_{\epsilon}$
(
$
\in \tau{{}_X} $)
such that
it holds that
$
|
L( x, \rho )
-
\widehat{L}(\Xi
, \rho )|
<
\epsilon
$
for
any open set
$\Xi$
$(
\in \tau{{}_X}
)$
satisfying
$x \in \Xi \subseteq \Xi_\epsilon$.
%
\par
\noindent
%
\rm
Let
${\mathcal N}_H$
be 
as in (F).
And consider a measurement
${\mathsf M}_{{\cal A}}({\mathsf O} $
$\equiv(X, {\cal F}, F{}) , S_{[\ast]})$).
Here define the
function
$\Lambda_{{\mathcal N}_H}{}: X \to [0,1]$
such that:
\begin{align*}
\Lambda_{{\mathcal N}_H}({}x)
=
\sup_{ \rho \in {\mathcal N}_H}
L(x, \rho )
\quad
({}\forall x \in X{}).
\tag{10}
\end{align*}
Also, for any
$\epsilon \; ({}0 < \epsilon \le 1{})$,
define
${{D}}_{{\mathcal N}_H}^\epsilon $
$({}\in \overline{\cal F}{})$
such that
\begin{align*}
{{D}}_{{\mathcal N}_H}^\epsilon
=
\{ x \in X \; | \;
\Lambda_{{\mathcal N}_H} ({}x{}) \le \epsilon \}.
\tag{11}
\end{align*}
\par
\noindent
\unitlength=0.20mm
\begin{picture}(400,130)
\put(27,18){0}
\put(27,50){$\epsilon$}
\put(27,108){1}
\put(350,18){$X$}
\dottedline{3}(40,110)(340,110)
\put(40,20){\line(0,1){100}}
\thicklines
\put(40,20){\line(1,0){300}}
\thicklines
\spline(40,70)(60,95)(75,108)(80,110)(85,110)(95,110)(100,110)
(150,100)(200,60)(250,40)
(270,35)(280,30)(300,25)(340,20)
\dottedline{5}(40,50)(340,50)
\dottedline{5}(225,50)(225,20)
\multiput(225,18)(3,0){40}{\line(0,2){4}}
\put(280,2){${{D}}_{{\mathcal N}_H}^\epsilon $}
\put(190,80){$\Lambda_{{\mathcal N}_H} (x) $}
\put(80,-20){\bf Figure 5.
\rm
$D_{{\mathcal N}_H}^\epsilon$
}
\end{picture}
\vskip0.5cm

\par
\noindent
Consider a positive number
$\alpha$
(called
\it
a significance level
\rm
)
such that
$0 < \alpha \ll 1$
({}e.g.
$\alpha = 0.05 $
).
\rm
Thus we can define
$\epsilon(\alpha)$
such that:
\begin{align*}
&
\epsilon(\alpha)
\\
=
&
\sup
\{
\epsilon \; |
\;
\sup_{ \rho \in {\mathcal N}_H }
{{}_{{\cal A}^*}} \langle 
\rho, 
F(
{{D}}_{{\mathcal N}_H}^\epsilon{}
)
\rangle_{{}_{\cal A}}
\le \alpha 
\}.
\tag{12} \end{align*}
It is clear that
the
${{D}}_{{\mathcal N}_H}^{\epsilon{(\alpha)}}$
satisfies the condition (G).
Then,
the rejection region
${\widehat R}^\alpha_{{\mathcal N}_H}$
is given by
${{D}}_{{\mathcal N}_H}^{{\epsilon(\alpha)} }$.

\par
\par
\par
\noindent
{\bf Remark 5.}
Note that
Problem (M$_1$) is not yet answered.
However,
we want to present the following conjecture:
\begin{itemize}
\item[(M$_2$)]
Under the general situation mentioned in
Section 4.1,
the
likelihood ratio test
(mentioned in Section 4.2)
is the only
statistical hypothesis testing.
Thai is, it is best.
\end{itemize}
The reason that we think so
is
that
we can not 
come up with another proper idea,
since our situation in Section 3.1
is too general.

\vskip0.5cm

\par
\par
\noindent
\par
\rm

\noindent
\subsection{\normalsize
Typical Examples in Classical Measurements
}

\rm

%

\par
\noindent
\sf
\rm
Our argument in the previous section
may be too abstract and general.
However, it is surely usual.
In this section,
this will be shown as easy examples
in classical measurements.

Put
$\Omega = {\mathbb R}$,
${\cal A} = {C} ({}\Omega{})$.
Fix $\sigma >0$.
And consider the normal observable
${\mathsf O}_{\sigma} $
$\equiv$
$({}{\mathbb R}{} , {\cal B}_{{\mathbb R}{}}^{} , F_{\sigma}{})$
in
${C} ({}\Omega{})$ such that:
\par
\noindent
\begin{align*}
&
[F_{\sigma}({\Xi})] ({} {}{\omega} {}) =
\frac{1}{{\sqrt{2 \pi }\sigma{}}}
\int_{{\Xi}} \exp[{}- \frac{({}{}{x} - {}{\omega}  {})^2 }{2 \sigma^2}    {}] d {}{x}
\\
&
\quad
({}\forall  {\Xi} \in {\cal B}_{{\mathbb R}{}}^{},
\quad
\forall   {}{\omega}    \in \Omega = {\mathbb R}{}).
\tag{13}
\end{align*}
And further, consider the product observable
${\mathsf O}_{\sigma}^2 $
$\equiv$
$({}{\mathbb R}^2{} , {\cal B}_{{\mathbb R}^2{}}^{} ,
F_{\sigma}^2 )$
in
${C} ({}\Omega{})$.
That is,
\par
\noindent
\begin{align*}
&
[F_{\sigma}^2
(\Xi_1 \bigtimes \Xi_2)]
({}\omega{})
=
[F_{\sigma}(\Xi_1)](\omega) \cdot [F_{\sigma}(\Xi_2)]
({}\omega{})
\\
=
&
\frac{1}{({{\sqrt{2 \pi }\sigma{}}})^2}
\iint_{\Xi_1 \times \Xi_2 }
\exp[{}- \frac{\sum_{k=1}^2 ({}{}{x_k} - {}{\omega}  {})^2 
}
{2 \sigma^2}    {}] d {}{x_1} d {}{x_2}
\\
&
\qquad 
({}\forall  \Xi_k \in {\cal B}_{{\mathbb R}{}}^{}
({}k=1,2),
\quad
\forall   {}{\omega}    \in \Omega = {\mathbb R}{}).
\tag{14}
\end{align*}
In what follows, we consider the
measurement
${\mathsf M}_{{C(\Omega)}}({\mathsf O}_\sigma^2 
=({}{\mathbb R}^2{} , {\cal B}_{{\mathbb R}^2{}}^{} ,
F_{\sigma}^2 )
,$
$ S_{[\ast]})$).
\par

\par
\noindent
{\bf
[Case(I)}; Two sided test, i.e., ${\mathcal N}_H = \{ \omega_0 \}${}].
Assume that
${\mathcal N}_H = \{ \omega_0 
\}$,
$\omega_0 \in \Omega = {\mathbb R}$.
Note the identification
(1),
i.e.,
$\delta_{\omega_0} \approx \omega_0$.
Then,
we see
that,
for any
$ {({}x_1 , x_2{})} \in {\mathbb R}^2{}$,
\begin{align*}
&
\; \;
\Lambda_{{\mathcal N}_H}({}x_1, x_2{})
=
\sup_{\omega \in \{ \omega_0 \} }
L( (x_1. x_2 ), \delta_{\omega} )
\\
&
=
\lim_{ {\Xi_1 \times \Xi_2} \to  ({}x_1 , x_2{})  }
\frac{
[F_{\sigma}^2
(\Xi_1 \bigtimes \Xi_2)]
({}\omega_0{})
}
{\sup_{ \omega \in \Omega }
[F_{\sigma}^2
(\Xi_1 \bigtimes \Xi_2)]
({}\omega{})
}
\\
&
=
\frac{
\exp[{}- \frac{({}{}{x_1} - {}{\omega_0}  {})^2 + ({}{}{x_2} - {}{\omega_0}  {})^2}
{2 \sigma^2}    {}]}
{
\exp[{}- \frac{({}{}{x_1} - {}{(x_1+x_2)/2}  {})^2 +
({}{}{x_2} - {}{(x_1+x_2)/2}  {})^2}
{2 \sigma^2}    {}]
}
\\
&
=
\exp[{}- \frac{[{}({}x_1 + x_2{})- 2 \omega_0]^2}
{4 \sigma^2}    {}].
\tag{15}
\end{align*}
Also, for any
$\epsilon ({}> 0{})$,
define
${{D}}_{\{ \omega_0 \}}^\epsilon $
$({}\in {\cal B}_{{\mathbb R}^2} {})$
such that:
\begin{align*}
{{D}}_{\{ \omega_0 \} }^\epsilon
=
\{ {({}x_1 , x_2{})} \in {\mathbb R}^2 \; | \;
\Lambda_{\{ \omega_0 \}} ({} x_1 , x_2 {}) \le \epsilon \}.
\tag{16} \end{align*}
Thus we can define
$\epsilon(\alpha)$
such that:
\begin{align*}
\epsilon(\alpha)
=
\sup
\{
\epsilon \; |
\;
\sup_{ \omega \in \{ \omega_0 \} }
[F_\sigma^2 ({{D}}_{\{ \omega_0 \}}^\epsilon{})](\omega)
\le \alpha
\}.
\tag{17}
\end{align*}
Thus, putting $\alpha = 0.05$,
we see that
\begin{align*}
&
{\widehat R}^{0.05}_{\{ \omega_0 \}}
=
{{D}}^{\epsilon(0.05)}_{\{ \omega_0 \} }
\\
=
&
\{ ({}x_1 , x_2{})
\in {\mathbb R}^2 \; \; |\; \;
({}x_1 + x_2{})/ 2 \le \omega_0 - 1.96  \sigma/ {\sqrt 2} \}
\\
&
\;
\bigcup
\{ ({}x_1 , x_2{})
\in {\mathbb R}^2  \; |\; 
({}x_1 + x_2{})/ 2 \ge \omega_0 + 1.96  \sigma/ {\sqrt 2} \}
\\
=
&
\text{{\lq\lq}Slash part in {\bf Figure 6}{\rq\rq}}
\tag{18} 
\end{align*}
\par
\noindent
\par
\noindent
\unitlength=0.4mm
\begin{picture}(300,150)
\put(70,50)
{{
\put(120,-10){$x_1$}
\put(-10,90){$x_2$}
\put(-30,0){\vector(1,0){160}}
\put(0,-20){\vector(0,1){120}}
\multiput(-35,50)(3,-3){20}{\line(-1,-1){30}}
\multiput(-5,70)(3,-3){32}{\line(1,1){30}}
\path(15,-2)(15,2)
\path(40,-2)(40,2)
\path(-2,15)(2,15)
\path(65,-2)(65,2)
\path(-2,65)(2,65)
\path(-2,40)(2,40)
\put(2,37){$2 \omega_0$}
\put(35,-8){$2 \omega_0$}
\put(17,3){$a$}
\put(62,-8){$b$}
\put(3,13){$a$}
\put(-6,62){$b$}
\put(40,94){ $a=2(\omega_0 - 1.96  \sigma/ {\sqrt 2})$}
\put(40,81){ $b=2(\omega_0 + 1.96  \sigma/ {\sqrt 2})$}
\thicklines
\put(-55,70){\line(1,-1){90}}
\put(-15,80){\line(1,-1){105}}
}}
\put(40,5){\bf Figure 6.
\rm
Rejection region
${\widehat{R}}^{0.05}_{\{ \omega_0 \}}$
}
%
\end{picture}
\par

\par
\noindent
{\bf [Case(II)}; One sided test, i.e., ${\mathcal N}_H = [\omega_0 , \infty{}) ${}].
Assume that
${\mathcal N}_H = [\omega_0 , \infty{})$,
$\omega_0 \in \Omega = {\mathbb R}$.
Then,
\begin{align*}
&
\; \;
\Lambda_{[{}\omega_0, \infty{}) }({} x_1 , x_2{})
=
\sup_{\omega \in [\omega_0, \infty) }
L( (x_1. x_2 ), \delta_{\omega} )
\\
&
=
\underset{{ \omega \in [\omega_0,\infty) }}{\sup}
\lim_{ {\Xi_1 \times \Xi_2} \to  ({}x_1 , x_2{})  }
\frac{
[F_{\sigma}^2
(\Xi_1 \bigtimes \Xi_2)]
({}\omega{})
}
{
\underset{{ \omega \in \Omega }}{\sup}
[F_{\sigma}^2
(\Xi_1 \bigtimes \Xi_2)]
({}\omega{})}
\\
&
=
\underset{{ \omega \in [\omega_0,\infty) }}{\sup}
\exp[{}- \frac{[{}({}x_1 + x_2{})- 2 \omega]^2}
{4 \sigma^2}    {}]
\\
&
=
\cases
\exp[{}- \frac{[{}({}x_1 + x_2{})- 2 \omega_0]^2}
{4 \sigma^2}    {}]
\quad &
({}\frac{x_1 + x_2}{2} < \omega_0 {})
\\
1
\quad &
(\text{ otherwise }{})
\endcases
\tag{19}
\end{align*}

\par
\noindent
Also, for any
$\epsilon ({}> 0{})$,
define
${{D}}_{ { [\omega_0, \infty{})  }}^\epsilon $
$({}\in {\cal B}_{{\mathbb R}^2}{})$
such that:
\begin{align*}
&
\;\;
{{D}}_{ { [{}\omega_0, \infty{})  }}^\epsilon
\\
&=
\{ ({}x_1 , x_2{}) \in {\mathbb R}^2 \; | \;
\Lambda_{{ [{}\omega_0, \infty{})  }} ({} x_1 , x_2 {}) \le \epsilon \}
\\
&
=
\{ ({}x_1 , x_2{}) \in {\mathbb R}^2 \; | \;
\frac{ x_1 + x_2 }{2} - \omega_0< {\sqrt{ 4\sigma^2 \log \epsilon }} \}.
\tag{20} \end{align*}
\par
\noindent
Thus we can define
$\epsilon(\alpha)$
such that:
\begin{align*}
\epsilon(\alpha)
=
\sup
\{
\epsilon \; |
\;
\sup_{ \omega \in { [{}\omega_0, \infty{})  } }
[F_{\sigma}^2
({{D}}_{ { [{}\omega_0, \infty{})  }}^\epsilon{})
]
({}\omega{})
\le
\alpha
\}.
\tag{21} \end{align*}
\par
\noindent
Therefore, putting
$\alpha = 0.05$,
we see that
\begin{align*}
&
{\widehat{R}}^{0.05}_{[\omega_0, \infty)}
=
{{D}}^{\epsilon(0.05)}_{[\omega_0,\infty{}) }
\\
=
&
\{ ({}x_1 , x_2{}) \in {\mathbb R}^2 \; \; |\; \;
({}x_1 + x_2{})/ 2 \le \omega_0 - 1.65  \sigma/ {\sqrt 2} \}
\\
=
&
\text{{\lq\lq}Slash part in \textcolor{black}{\bf Figure 7}{\rq\rq}}
\tag{22} 
\end{align*}
\par
\noindent
\par
\noindent
\noindent

\par
\noindent
\unitlength=0.43mm
\begin{picture}(300,120)
\put(75,50)
{{
\put(100,-10){$x_1$}
\put(-10,65){$x_2$}
\put(-30,0){\vector(1,0){130}}
\put(0,-20){\vector(0,1){90}}
\multiput(-35,50)(3,-3){20}{\line(-1,-1){30}}
\path(15,-2)(15,2)
\path(40,-2)(40,2)
\path(-2,15)(2,15)
\path(-2,40)(2,40)
\put(2,37){$2 \omega_0$}
\put(35,-8){$2 \omega_0$}
\put(17,3){$c$}
\put(3,13){$c$}
\put(25,50){ $c=2(\omega_0 - 1.65  \sigma/ {\sqrt 2})$}
\thicklines
\put(-50,65){\line(1,-1){85}}
}}
\put(40,5){\bf Figure 7.
\rm
Rejection region
${\widehat{R}}^{0.05}_{[\omega_0, \infty)}$
}
\end{picture}

%
%
%
%
%
%

\par
\par
\noindent
{\bf [Case(III)}; ${\mathcal N}_H ={\mathbb Q}$,
i.e.,
the set of all rational numbers].
It is clear that
$\Lambda_{\mathbb Q} (x_1, x_2) =1$,
$( \forall (x_1, x_2 ) \in {\mathbb R}^2 )$.
Thus, the rejection region does not exist.

\par
\par
\noindent
\section{
\large
\bf
Fisher-Bayes Method in classical
MT
}

\par
\par
\noindent
\par

\par

\par
\par

\subsection{\normalsize
Bayes Method
in Classical
$C(\Omega)$
}

\par
\noindent

\rm 
Let ${\mathsf O}_1 \equiv (X, {\cal F}, F)$ be an observable in a commutative $C^*$-algebra $C(\Omega)$.
And let
${\mathsf O}_2 \equiv (Y, {\cal G}, G)$ be any observable in $C(\Omega)$.
Consider the product observable
${\mathsf O}_1 \times {\mathsf O}_2 \equiv (X\times Y, {\cal F}
\bigstimes
{\cal G}, F \times G)$ in $C(\Omega)$.
The existence will be shown
in Section 7 in \cite{Ishi7}.

Assume that
we know that
the measured value $(x,y)$ obtained by a simultaneous measurement
${\mathsf M}_{C(\Omega)}(
{\mathsf O}_1 \times {\mathsf O}_2,
S_{[*]}
{(\{\nu_0\})}
)$
belongs to
$\Xi \times Y \;(\in {\cal F} \boxtimes {\cal G} )$.
Then,
by \textcolor{black}{(5)},
we can infer that
\begin{itemize}
\item[{\rm (N)}]
the probability $
P_\Xi (G(\Gamma))
$
that
$y$ belongs to $\Gamma (\in {\cal G})$
is given by
$$
\!\!\!
P_\Xi (G(\Gamma))
= \frac{\int_{\Omega} [F(\Xi)G(\Gamma)](\omega)
\;
\nu_0 (d \omega) }{
\int_{\Omega} [F(\Xi)](\omega)
\;
\nu_0(d \omega )}
\;\;
(\forall \Gamma \in {\cal G}).
$$
\end{itemize}
Thus,
we can assert that:

\par
\noindent
{\bf Theorem 2}
\rm
[Bayes method,
{\rm cf.$\;$}\textcolor{black}{\cite{Ishi3,Ishi4, Ishi7}}].
\it
When
we
know that
a measured value
obtained by
a
measurement
${\mathsf M}_{C(\Omega)}(
{\mathsf O}_1 \equiv (X, {\cal F}, F)
, S_{[*]}
{(\{ \nu_0\})}
)$
belongs to
$\Xi$,
there is a reason to
infer
that
the mixed state after the measurement
is
equal to
$\nu_0^a$
$( \in {\cal M}_{+1}^m (\Omega ))$,
where
$$
\nu_0^a( D)= \frac{\int_{D} [F(\Xi)](\omega)
\;
\nu_0 (d \omega) }{
\int_{\Omega} [F(\Xi)](\omega)
\;
\nu_0(d \omega )}
\quad
(\forall D \in {\cal B}_\Omega ).
$$
\par
\noindent
{\it Proof.}
\rm
Note that
we can regard that
$P_\Xi$
$\in {\cal M}_{+1}^m (\Omega ) ( \subseteq C(\Omega )^*)$.
That is,
there exists
$\nu_0^a$
$(
\in
{{C(\Omega)}^*}
)$
such that
\begin{align*}
P_\Xi (G(\Gamma)
)
=
\int_\Omega
[G(\Gamma)
]
( \omega ) \;
\nu_0^a (d \omega )
\quad
(\forall \Gamma \in {\cal G})
\tag{23}
\end{align*}
Then,
Axiom$^{\rm S}$1 says that
the probability that a measured value $y$
$( \in Y)$ obtained by the measurement 
${\mathsf{M}}_{{{C(\Omega)}}} ({\mathsf{O}}_2$
${ \equiv} (Y, {\cal G}, G),$
{}{$ S_{[\ast]}(\{ \nu_0^a \}) )$}
belongs to a set 
$\Gamma (\in {\cal G})$ is given by
$
\int_\Omega [G(\Gamma)] (\omega) \; \nu_0^a (d \omega )
$,
which is equal to
$P_\Xi (G(\Gamma))$
in \textcolor{black}{(23)}.
Since
${\mathsf O}_2 \equiv (Y, {\cal G}, G)$
is
arbitrary,
we obtain \textcolor{black}{Theorem 2}.

\rm
\par
\vskip0.3cm
\par
\noindent
{\bf Remark 6.}
The above \textcolor{black}{(N)} is,
of course, fundamental.
However,
in the sense mentioned in the above proof,
we admit Theorem 2
as the equivalent statement
of the \textcolor{black}{(N)}.
That is, in spite of Interpretation (F$_2$),
we admit
the wavefunction collapse such as
\begin{itemize}
\item[(O$_1$)]
$
\quad
\overset{\text{(pretest state)}}{
\underset{
(\in {\cal M}_{+1}^m(\Omega))}
{\nu_0}
}
\xrightarrow[\text{\tiny Theorem 2}]{\text{\tiny Bayes}}
\overset{\text{(posttest state)}}{
\underset{(
\in {\cal M}_{+1}^m(\Omega))}
{\nu_0^a}
}
$
\end{itemize}
Theorem 2 was, for the first time, proposed 
in \textcolor{black}{\cite{Ishi3, Ishi4}}
without
the conscious understanding of
Interpretation (F$_2$).
Also, note that,
\begin{itemize}
\item[(O$_2$)]
in Theorem 2,
if $\nu_0= \delta_{\omega_0}$
$( \in {\cal M}_{+1}^p(\Omega ) )$,
then it clearly holds that
$\nu_0^a= \delta_{\omega_0}$.
\end{itemize}
Also,
for our opinion concerning the wavefunction collapse
in quantum mechanics,
see \textcolor{black}{\cite{Ishi5}}.

\par
\par
\noindent
\par
\noindent
\subsection{\normalsize
Fisher-Bayes Method
in Classical
$C(\Omega)$
}
\rm

Combining Theorem 1 (Fisher's method) and Theorem 2 (Bayes' method),
we get the following corollary.

\par
\noindent
{\bf Corollary 2}
\rm
[Fisher-Bayes method
(i.e.,
Regression analysis in a narrow sense)
].
\it
When
we
know that
a measured value
obtained by
a
measurement
${\mathsf M}_{C(\Omega)}(
{\mathsf O}_1 \equiv (X, {\cal F}, F)
, S_{[*]}
(K{)}
)$
belongs to
$\Xi$,
there is a reason to
infer that
the state after the measurement
is
equal to
$\nu_0^a$
$( \in {\cal M}_{+1}^m (\Omega ))$
such that
$$
\nu_0^a( D)= \frac{\int_{D} [F(\Xi)](\omega)
\nu_0 (d \omega) }{
\int_{\Omega} [F(\Xi)](\omega)
\nu_0(d \omega )}
\quad
(\forall D \in {\cal B}_\Omega )
$$
where
the $\nu_0
(\in K)$
is defined by
$$
\int_\Omega [F(\Xi)](\omega)
\;
\nu_0(d \omega )
= \max_{\nu \in K } 
\int_\Omega [F(\Xi)](\omega)\nu(d \omega ).
$$
\rm
\rm
\par
\vskip0.3cm
\par
\par
\noindent
{\bf Remark 7.}
As mentioned in the above,
note that
Corollary 2
is composed of the following two procedure:
\begin{itemize}
\item[(O$_3$)]
$
\underset{(\subseteq {\cal M}_{+1}^m(\Omega))}{
K}
\xrightarrow[\text{\tiny Theorem 1}]{\text{\tiny Fisher}}
\underset{
(\in 
K
)}
{\nu_0}
\xrightarrow[\text{\tiny Theorem 2}]{\text{\tiny Bayes}}
\underset{(
\in 
K
)
}
{\nu_0^a}
$
\end{itemize}

\par
\noindent

\par
\par

\subsection{\normalsize
A Simple example of Fisher-Bayes Method
(Regression Analysis in a Narrow Sense)
}

\par
\noindent
In this section,
we examine Corollary 2
in a simple example.
Readers will find that
Corollary 2 can be regarded as regression analysis
in a narrow sense.

\rm
We have a rectangular water tank filled with water.
Assume that
the height of water at time $t$ is given
by the following function $h(t)$:
\begin{align*}
h(t) = \alpha_0 + \beta_0 t,
\tag{24}
\end{align*}
where
$\alpha_0$
and
$\beta_0$
are unknown fixed parameters
such that
$\alpha_0$ is the height of water filling the tank at the beginning and $\beta_0$ is the increasing height of water per unit time.
The measured height $h_m(t)$ of water at time $t$
is assumed to be represented by
\begin{align*}
h_m(t) = \alpha_0 + \beta_0 t + e(t),
\tag{25}
\end{align*}
where $e(t)$ represents a noise
(or more precisely,
a measurement error) with some suitable conditions.
And assume that
we obtained the measured data of the heights of water
at $t=0,1,2$ as follows:
\begin{align*}
h_m(0)=0.5, \quad h_m(1)=1.6, \quad h_m(2)=3.3.
\tag{26}
\end{align*}
\par
\noindent
\unitlength=0.30mm
\begin{picture}(200,130)
\put(-70,0){
\put(100,0){
\path(130,112)(130,70)
\path(133,112)(133,70)
\path(136,112)(136,70)
\path(139,112)(139,70)
\put(165,30){$h(t)$}
\put(168,25){\vector(0,-1){15}}
\put(168,35){\vector(0,1){22}}
\thicklines
\put(10,10){\line(0,1){100}}
\put(10,10){\line(1,0){150}}
\path(160,10)(160,110)
\path(180,120)(140,120)(140,110)
\path(129,110)(129,130)(180,130)
\multiput(10,10)(0,3){17}{\line(1,0){150}}
}
}
\put(40,-4){\bf Figure 8.
\rm
Water tank
}
\end{picture}

\vskip0.5cm

\par
\noindent
Under this setting, we shall study the following problem:
\begin{itemize}
\item[(P)]
[Inference]:
when measured data
\textcolor{black}{(26)}
is obtained,
infer the unknown parameter $(\alpha_0, \beta_0)$
in \textcolor{black}{(25)}.
\end{itemize}
In what follows,
from the measurement theoretical point of view,
we shall answer the problem \textcolor{black}{(P)}.
Let
$T=\{0,1,2\}$
be 
a series ordered set
such that
the parent map $\pi :T\setminus\{0\} \to T$
is defined by
$\pi ( t ) = t-1$
$\;(t=0,1,2)$.
Put
$\Omega_0
= [0,\; 2] \times [0,\; 2]$,
$\Omega_1 =[0,\; 4] \times [0,\; 2]$,
$\Omega_2 = [0,\; 6] \times [0,\; 2]$.
For
each
$t=1,2$,
consider
a continuous map
$\phi_{\pi(t),t}{}: \Omega_{\pi(t)} \to \Omega_t $
such that
\begin{align*}
&
\phi_{0,1}(\alpha , \beta ) = (\alpha + \beta, \beta)
&\;&
(\forall \omega_0 =(\alpha, \beta ) \in \Omega_0 )
\\
&
\phi_{1,2}(\alpha , \beta ) = ( \alpha + \beta, \beta)
&\;&
(\forall \omega_1 =
(\alpha, \beta ) 
\in \Omega_1 
).
\tag{27} \end{align*}
Then,
we get
the deterministic causal operators
hus,
$\{\Phi_{\pi(t),t}{}: C(\Omega_{t}) \to C(\Omega_{\pi(t)}
)
\}_{t \in \{1,2\}}$
such that
\begin{align*}
(\Phi_{0,1}f_1)(\omega_0) \!=\! f_1( \phi_{0,1}(\omega_0))
&
\;
(\forall f_1 \in C(\Omega_1),
\forall \omega_0 \in \Omega_0)
\\
(\Phi_{1,2}f_2)(\omega_1) \!=\! f_2( \phi_{1,2}(\omega_1))
&
\;
(
\forall f_2 \in C(\Omega_2),
\forall \omega_1 \in \Omega_1).
\tag{28}
\end{align*}
Thus,
we have the causal relation as follows.
\begin{align*}
{
{\text{$ {C (\Omega_0)} $}}
}
\mathop{\longleftarrow}^{\Phi_{0, 1 } }
{\text{$ {C (\Omega_{1})} $}}
\mathop{\longleftarrow}^{\Phi_{1,2 } }
{\text{$ {C (\Omega_{2})} $}}.
\end{align*}
Put
$\phi_{0,2}(\omega_0)=\phi_{1,2}(\phi_{0,1}(\omega_0))$,
$\Phi_{0,2}=\Phi_{0,1}\cdot \Phi_{1,2}$.

Let ${\mathbb R}$ be the set of real numbers.
Fix
$\sigma>0$.
For each
$t=0,1,2$,
define
the {\it
normal observable}
${\mathsf O}_{t} {{\equiv}} ({\mathbb R}, {\cal B}_{\mathbb R}, G^n_{\sigma})$
in
$C (\Omega_t)$
such that
\begin{align*}
&
[G^n_{\sigma}(\Xi)] (\omega_t ) = \frac{1}{\sqrt{2 \pi \sigma^2}}
\int_{\Xi} \exp({- \frac{(x - \alpha)^2}{2 \sigma^2}}) dx
\\
&
(\forall \Xi \in {\cal B}_{\mathbb R}, \forall \omega_t
=(\alpha, \beta) \in \Omega_t
{{=}}
[{}0, \; 2t+2{}]\times [0, \;2]).
\tag{29} \end{align*}
Thus,
we get
the
sequential deterministic causal observable
$[{\mathbb O}_{T{}}]$
${{=}}$
$[ \{{\mathsf O}_{t}\}_{t=0,1,2}
,
\{\Phi_{\pi(t),t}{}: C(\Omega_{t}) \to C(\Omega_{\pi(t)}
)
\}_{t=1,2}
]$.
Then,
the
realized causal observable
$\widehat{\mathsf O}_{0{}}$
${{\equiv}}$
$({\mathbb R}^3, {\cal B}_{{\mathbb R}^3}, {\widehat F}_0)$
in
${C(\Omega_0{})}$
is,
by (4) and (28),
obtained
as follows:
\begin{align*}
&
[{\widehat F}_0({}\Xi_0 \times \Xi_1 \times \Xi_2{})]
(\omega_0)
\\
=
&
\big[
\big(G^n_{\sigma} ({\Xi_0})
\Phi_{0,1} (G^n_{\sigma} ({\Xi_1})
\Phi_{1,2} (G^n_{\sigma} ({\Xi_2})
))
\big)
\big]
(\omega_0)
\\
=
&
[G^n_{\sigma} ({\Xi_0})] ({}\omega_0{})
\cdot
[G^n_{\sigma} ({\Xi_1})] ({}\phi_{0,1}(\omega_0){})
\\
&
\qquad \qquad
\cdot
[G^n_{\sigma}({\Xi_2})] ({}\phi_{0,2}(\omega_0){})
%
\\
&
({}\forall \Xi_0, \Xi_1, \Xi_2 \in {\cal B}_{\mathbb R},
\;
\forall \omega_0 =({}\alpha, \beta{}) \in \Omega_0
{}).
\tag{30} \end{align*}
Putting
$K={\cal M}_{+1}^p(\Omega_0)$,
we have the measurement
${\mathsf M}_{C({}\Omega_0{})} (\widehat{\mathsf O}_0,$
$
S_{[\ast]}{}
(
{\cal M}^p_{+1} (\Omega_0)
)\;)$.
Recall the
\textcolor{black}{(26)},
that is,
the measured value
$(x_0, x_1, x_2 )$
obtained by
the measurement
${\mathsf M}_{C({}\Omega_0{})} (\widehat{\mathsf O}_0,$
$
S_{[\ast]}{}
(
{\cal M}^p_{+1} (\Omega_0)
)\;)$
is equal to
\par
\begin{align*}
(0.5, \; 1.6, \; 3.3) \; (\in {\mathbb R}^3).
\tag{31}
\end{align*}
\par
\noindent
Define the closed interval
$\Xi_t$
$(t=0,2,3)$
such that
\begin{align*}
&
\Xi_0 =[{}0.5 - \frac1{2N}, 0.5 + \frac1{2N}],
\\
&
\Xi_1 =[{}1.6 - \frac1{2N}, 1.6 + \frac1{2N}],
\\
&
\Xi_2 =[{}3.3 - \frac1{2N}, 3.3 + \frac1{2N}],
\end{align*}
for sufficiently large $N$.
Here,
Fisher's method
(\textcolor{black}{Theorem 1}) says that
it suffices to
solve the problem.
\begin{itemize}
\item[(Q$_1$)]
Find
$({}\alpha_0, \beta_0{})$
such as
\begin{align*}
\max_{ ({}\alpha, \beta{}) \in \Omega_0 }
[{\widehat F}_0(
\Xi_0
\times
\Xi_1
\times 
\Xi_2
]
({}\alpha, \beta{})
\tag{32}
\end{align*}
\end{itemize}
\par
\noindent

Putting
\begin{align*}
&
U(x_0, x_1,x_2, \alpha, \beta)
=
\sum_{k=0}^2
({}{}{x_k} - ({}\alpha + k \beta{}){})^2 
\end{align*}
we have the following problem
that is equivalent to
\textcolor{black}{(O$_1$)}:
\begin{itemize}
\item[(Q$_2$)]
Find
$({}\alpha_0, \beta_0{})$
such as
\begin{align*}
&
\min_{ ({}\alpha, \beta{}) \in \Omega_0 }
\exp \Big(- \frac{U(x_0, x_1,x_2, \alpha, \beta)}{2 \sigma^2}
\Big)
\\
\Leftrightarrow
&
\max_{ ({}\alpha, \beta{}) \in \Omega_0 }
U(x_0, x_1,x_2, \alpha, \beta).
\end{align*}
\end{itemize}
Calculating
\begin{align*}
&
\frac{\partial}{\partial \alpha}{U(0.5,1.6,3.3, \alpha,\beta) }
=0,
\\
&
\frac{\partial}{\partial \beta}{U(0.5,1.6, 3.3, \alpha,\beta)}
=0,
\end{align*}
we get
\begin{align*}
({}\alpha, \beta{}) =
(0.4, 1.4)
\tag{33} 
\end{align*}
Thus, we see, by the statement \textcolor{black}{(O$_2$)}, that
\begin{itemize}
\item[(R)]
$
\underset{(\subseteq {\cal M}_{+1}^m(\Omega))}{
{\cal M}_{+1}^p(\Omega_0)}
\xrightarrow[\text{\tiny Theorem 1}]{\text{\tiny Fisher}}
\underset{
(\in K)}
{\delta_{(0.4, 1.4)}}
$
\\
$\qquad \quad$
$\qquad \qquad$
$
\xrightarrow[\text{\tiny Theorem 2}]{\text{\tiny Bayes}}
\underset{(
\in K)}
{\delta_{(0.4, 1.4)}}
$
\end{itemize}
This (i.e.,
$(\alpha_0, \beta_0)=
(0.4, 1.4)$
)
is the answer to the problem \textcolor{black}{(P)}.

\par
\noindent
\bf
Problem 1.
\rm
Since the above example is quite easy,
the validity of Bayes' theorem
in (R)
may not be clear.
If it is so,
instead of
the problem \textcolor{black}{(O$_3$)},
we should present
the following simple problem.
\begin{itemize}
\item[(S)]
Infer the water level at time $1$.
\end{itemize}
Some may calculate and conclude as follows:
\begin{align*}
h(1)=\alpha_0 + \beta_0 \times 1 =0.4  + 1.4 = 1.8
\tag{34}
\end{align*}
However,
this calculation is based on
the Schr\"{o}dinger picture,
and thus,
the justification of this calculation \textcolor{black}{(35)} is not assured.
That is because
measurement theory
(particularly,
Interpretation \textcolor{black}{(F$_2$)}
)
says that
the Heisenberg picture should be adopted.
Therefore,
in order to answer the problem \textcolor{black}{(S)},
we must prepare Corollary 3
(i.e.,
regression analysis in a wide sense)
in the following section.

\par
\vskip0.3cm
\par
\noindent
{\bf Remark 8}.
\rm
It should be noted that the following two are equivalent:
\begin{itemize}
\item[(T$_1$)]
[=\textcolor{black}{(P)}; Inference]:
when measured data
\textcolor{black}{(26)}
is obtained,
infer the unknown parameter $(\alpha_0, \beta_0)$.
\end{itemize}
\begin{itemize}
\item[(T$_2$)]
[Control]:
Settle the parameter $(\alpha_0, \beta_0)$
such that
measured data
\textcolor{black}{(26)}
will be obtained.
\end{itemize}
That is,
we see that
$$
\text{{\lq\lq}inference"}=\text{{\lq\lq}control"}\!.
$$
Hence,
from the measurement theoretical point of view,
we consider that
\begin{itemize}
\item[]
$
\text{{\lq\lq}Statistics"}=\text{{\lq\lq}Dynamical system theory"}\!,
$
\end{itemize}
though
these are superficially different
in applications.

\rm


\par
\par
\noindent
\section{
\large
Causal Fisher-Bayes method in classical MT
}

\par
\par
\noindent

\subsection{\normalsize
Causal Bayes Method
in Classical
$C(\Omega)$
}

\par
\noindent

Let
$t_0$
be the root
of a tree $T$.
Let
$[{\mathbb O}_T^\times{}]$
$=$
$[{}
\{ {\mathsf{O}}_t^\times  ({}\equiv ({}X_t \times Y_t  ,$
$ {\cal F}_{t} \boxtimes 
{\cal G}_t , {F}_t \times G_t ))
\}_{ t \in T} ,
\{  \Phi_{t_1,t_2}{}: $
$C(\Omega_{t_2}) \to C(\Omega_{t_1}) \}_{(t_1,t_2) \in T^2_\le }$
$]$
be
a sequential causal observable
with
the realization
$\widehat{\mathsf{O}}_{t_0}^\times $
$\equiv$
$(\bigtimes_{t \in T } (X_t \times Y_t) , $
$\bigstimes_{t \in T }
({\cal F}_{t} \boxtimes {\cal G}_t),$
${\widehat H}_{t_0})$
in $C(\Omega_{t_0})$.
Thus we have the statistical measurement
${\mathsf M}_{C(\Omega_{t_0})}
(\widehat{\mathsf{O}}_{t_0}^\times, S_{[\ast]}(\{\nu_0 \}) )$,
where
$\nu_0 \in {\cal M}_{+1}^m (\Omega_{t_0} )$.
Assume that
we know that
the measured value $(x,y)$
$(
=
(
(x_t)_{t\in T},
(x_t)_{t\in T},
)
\in (\bigtimes_{t \in T} X_t)
\bigtimes
(\bigtimes_{t \in T} Y_t)
)$
obtained by the measurement
${\mathsf M}_{C(\Omega_{t_0})}
(\widehat{\mathsf{O}}_{t_0}^\times, S_{[\ast]}(\{\nu_0 \}) )$
belongs to
$(\bigtimes_{t \in T} \Xi_t)
\bigtimes 
(\bigtimes_{t \in T}Y_t) \;(\in 
(\boxtimes_{t \in T}{\cal F}_t)
\boxtimes 
(\boxtimes_{t \in T}Y_t)
)$.
Then,
by \textcolor{black}{(5)},
we can infer that
\begin{itemize}
\item[{\rm (U)}]
the probability $
P_{\times_{t \in T} \Xi_t} (
(G_t (\Gamma_t))_{t \in T}
)
$
that
$y$ belongs to $\bigtimes_{t \in T} \Gamma_t (\in \boxtimes_{t \in T} {\cal G}_t)$
is given by
\begin{align*}
&
P_{\times_{t \in T} \Xi_t} (
(G_t (\Gamma_t))_{t \in T}
)
\\
= 
&
\frac{\int_{\Omega} 
[
{\widehat H}_{t_0}
(
(\bigtimes_{t \in T} \Xi_t)
\bigtimes 
(\bigtimes_{t \in T} \Gamma_t) 
)
]
(\omega)
\nu_0 (d \omega) }{
\int_{\Omega} [
{\widehat H}_{t_0}
(\bigtimes_{t \in T} \Xi_t)
\bigtimes 
(\bigtimes_{t \in T}Y_t) 
](\omega)
\nu_0(d \omega )}
\\
&
\quad
(\forall \Gamma_t \in {\cal G}_t, t \in T).
\tag{35}
\end{align*}
\end{itemize}

Note that
we can regard that
$P_{\times_{t \in T} \Xi_t} $
$\in {\cal M}_{+1}^m (\bigtimes_{t \in T} \Omega_t )$
$( \subseteq C(\bigtimes_{t \in T} \Omega_t )^*)$.
That is,
there uniquely exists 
$
\nu_T^a
\in {\cal M}_{+1}^m (\bigtimes_{t \in T} \Omega_t )
$
such that
\begin{align*}
&
P_{\times_{t \in T} \Xi_t} (
(G_t (\Gamma_t))_{t \in T}
)
\\
=
&
\int_{\times_{t \in T}\Omega_t}
[
{\text{\footnotesize $\bigotimes$}}_{t\in T} G_t (\Gamma_t)
]
( \omega )
\; 
\nu_T^a (d \omega )
\tag{36}
\end{align*}
for any observable
$(Y_t, {\cal G}_t, G_t)$
in
$C(\Omega_t)$
$( t \in T)$.
Here, we used the following notation:
\begin{align*}
&
[
{\text{\footnotesize $\bigotimes$}}_{t\in T} G_t (\Gamma_t)
]
(\omega)
=
\bigtimes_{t \in T}
[G_t (\Gamma_t)](\omega_t)
\\
&
\qquad
\quad
(\forall \omega = (\omega_t)_{t \in T}
\in
\bigtimes_{t \in T} \Omega_t
).
\end{align*}
Define the observable
$\widehat{\mathsf{O}}_{t_0} $
$\equiv (\bigtimes_{t \in T } X_t , $
$\bigstimes_{t \in T } {\cal F}_{t} ,$
${\widehat F}_{t_0})$
such that
$$
{\widehat F}_{t_0}
(\bigtimes_{t \in T} \Xi_t)
=
{\widehat H}_{t_0}
(
(\bigtimes_{t \in T} \Xi_t)
\bigtimes 
(\bigtimes_{t \in T}Y_t)
).
$$
Then,
we can define the Bayes operator
$[B_{\widehat{\mathsf{O}}_{t_0} }(\bigtimes_{t \in T} \Xi_t)]:
{\cal M}_{+1}^m(\Omega_{t_0})$
$
\to {\cal M}_{+1}^m( \bigtimes_{t \in T}\Omega_t)$
by
\textcolor{black}{(36)}.

\par
\vskip0.3cm
\par
Thus,
as the generalization of Theorem 2,
we have:
\par
\noindent
\bf
Theorem 3
\rm
[Causal Bayes' theorem in classical measurements,
{\rm cf.$\;$}\textcolor{black}{\cite{Ishi7}}].
\it
Let
$t_0$
be the root
of a tree $T$.
Let
$[{\mathbb O}_T{}]$
$=$
$[{}
\{ {\mathsf{O}}_t ({}\equiv ({}X_t , {\cal F}_{t}, {F}_t ))
\}_{ t \in T} ,
\{  \Phi_{t_1,t_2}{}: $
$C(\Omega_{t_2}) \to C(\Omega_{t_1}) \}_{(t_1,t_2) \in T^2_\le }$
$]$
be
a sequential causal observable
with
the
realization
$\widehat{\mathsf{O}}_{t_0} $
$\equiv (\bigtimes_{t \in T } X_t , $
$\bigstimes_{t \in T } {\cal F}_{t} ,$
${\widehat F}_{t_0})$.
Thus we have the statistical measurement
${\mathsf M}_{C(\Omega_{t_0})}
(\widehat{\mathsf{O}}_{t_0}, S_{[\ast]}(\{ \nu_0 \}) )$,
where
$\nu_0 \in {\cal M}_{+1}^m (\Omega_{t_0}))$.
Assume that
we know that
a measured value obtained by
the statistical measurement
${\mathsf M}_{C(\Omega_{t_0})}
(\widehat{\mathsf{O}}_{t_0}, S_{[\ast]}(\{ \nu_0 \}) )$
belongs to
$\bigtimes_{t \in T} \Xi_t$.
Then,
there is a reason to infer that the mixed state
$\nu_T^{a} ( \in {\cal M}_{+1}^m(\bigtimes_{t \in T } \Omega_t ) )$
after the statistical measurement
${\mathsf M}_{C(\Omega_{t_0})}
(\widehat{\mathsf{O}}_{t_0}, S_{[\ast]}(\{ \nu_0 \}) )$
is given by
$[B_{\widehat{\mathsf{O}}_{t_0} }(\bigtimes_{t \in T} \Xi_t)]
(\nu_0)(
\in {\cal M}_{+1}^m (\bigtimes_{t \in T} \Omega_t )
)
$.

\par
\noindent
\par
\noindent
{\it Proof.}
\rm
The proof is similar to the proof of Theorem 2.
Thus, we omit it.
\qed

\rm
\par
\vskip0.3cm
\par
\noindent
{\bf Remark 9.}
In Theorem 3,
we see that
\begin{itemize}
\item[(V)]
$
\quad
\overset{\text{(pretest state)}}{
\underset{
(\in {\cal M}_{+1}^m(\Omega_{t_0}))}
{\nu_0}
}
\xrightarrow[\text{\tiny Theorem 2}]{\text{\tiny Bayes}}
\overset{\text{(posttest state)}}{
\underset{(
\in {\cal M}_{+1}^m(\times_{t\in T} \Omega_t))}
{\nu_T^a}
}
$
\end{itemize}
which is the generalization of the \textcolor{black}{(O$_1$)}.

\par
\rm
\vskip1.0cm
\rm

\par
The following example promotes the understanding of
Theorem 3.
\par
\noindent
\bf 
{Example 2}
\rm
[The simple case such that
$T=\{0,1,2\}$].
\rm
Consider a particular case
such that
$T= \{ 0,1,2 \}$
is series ordered set,
i.e.,
$\pi( t) = t-1$
$( \forall t \in T \setminus \{ 0 \} )$.
And consider a
causal relation
$\{
C(\Omega_t) {{\Phi_{ \pi({}t{}), t } }\atop{\rightarrow}} 
C(\Omega_{\pi({}t{})})
\}_{
t \in T \setminus \{ 0 \}
}$,
that is,
\begin{align*}
{
{\text{$ C(\Omega_0) $}}
}
\mathop{\longleftarrow}^{\Phi_{0, 1 } }
{\text{$ C(\Omega_1) $}}
\mathop{\longleftarrow}^{\Phi_{1,2 } }
{\text{$ C(\Omega_2) $}}.
\end{align*}
\par
\noindent
Further
consider sequential causal observable
$[{\mathbb O}_T]$
$=$
$[{}\{ {\mathsf O}_t \}_{ t \in T} ,
\{  \Phi_{t,\pi(t) }{}: $
$C(\Omega_{t}) \to C(\Omega_{\pi(t)}) \}_{ t \in T\setminus \{0\} }$
$]$.
Let
$
\widehat{\mathsf O}_0
\equiv
(\bigtimes_{t \in T}  X_t , \bigtimes_{t \in T}  {\Cal F}_t , 
{\widehat F}_0 )
$
be
its realization.
Note,
by the formula \textcolor{black}{(4)},
that,
\begin{align*}
&
{\widehat F}_0
(\Xi_0 \times
\Xi_1 \times
\Xi_2 
)
\\
=
&
\Phi_{0,1}
(F_0(\Xi_0)
(\Phi_{1,2}F_1(\Xi_1)(\Phi_{1,2}
(
F_2(\Xi_2)
))))
\\
&
\qquad
\qquad
(\Xi_t \in {\cal F}_t (\forall t \in T)).
\end{align*}
\rm
Putting 
$K=\{ \nu_0\}$,
we have the measurement
\begin{align*}
{\mathsf M}_{
C(\Omega_{0})
} ({}\widehat{\mathsf O}_0
\equiv ({\bigtimes}_{t \in T} X_t,
{{\bigstimes}_{t \in T} {\cal F}_t}
, {\widehat F}_0), S_{[\ast]} {(\{ \nu_0 \})}{}).
\tag{37} \end{align*}
Let
$\nu_T^a
(\in {\cal M}_{+1}^m( \Omega_0 \times \Omega_1 \times \Omega_2 )
$
be
the posttest state
in \textcolor{black}{(V)},
that is,
$\nu_T^a
$
$
=
$
$[B_{\widehat{\mathsf{O}}_{0} }(\bigtimes_{t \in T} \Xi_t)]
(\nu_0)$.
Define
$\nu_{\{1\}}^a
(\in {\cal M}_{+1}^m (\Omega_1 )\;)$
such that
$$
\nu_{\{1\}}^a (D_1 )
=
\nu_T^a
(
\Omega_0 \times D_1 \times \Omega_2
)
\quad
(\forall D_1 \in {\cal B}_{\Omega_1}
).
$$
Then,
we see that
\begin{align*}
&
\nu_{\{1\}}^a
=
\frac{ 
{}_{
{}_{} 
}
{ 
( F_{1}( \Xi_1 ) (\Phi_{{1}  , {{2} } } 
F_2 ( \Xi_2)  ) ( \Phi^*_{{0}  , {{1} } } ( F_0 ( \Xi_0 )   \nu_0 )) }
}
{ 
{
\Big\langle 
\nu_0,
F_0 ( \Xi_0 )  \Phi_{{0}  , {{1} } } \Big( F_{{1} } (\Xi_{{1} } ) \Phi_{{1}  , {{2} } }
\Big(   F_{{2} } (\Xi_{{2} } )    \Big) \Big) \Big\rangle  }
}.
\end{align*}
That is because
we see that,
for any observable
$(  Y_1 ,  {\Cal G}_{1} , $
$ G_1)$
in $C(\Omega_1)$,
\begin{align*}
&
\langle \nu_{\{1\}}^a , G_1 (\Gamma_1 )
\rangle
\\
=
&
\frac
{ 
{ \Big\langle  \nu_0,
F_0 ( \Xi_0 )  \Phi_{0  , 1} \Big( F_{1} (\Xi_{1} )  G_1( \Gamma_1 ) \Phi_{1,2 }
\Big(   F_{{2} } (\Xi_{{2} } )   \Big) \Big) \Big\rangle  }
}
{ 
{
\Big\langle 
\nu_0,
F_0 ( \Xi_0 )  \Phi_{{0}  , {{1} } } \Big( F_{{1} } (\Xi_{{1} } ) G_1 ( Y_1 )  \Phi_{{1}  ,
 {{2} } }
\Big(   F_{{2} } (\Xi_{{2} } )   \Big) \Big) \Big\rangle  }
}
\\
=
&
\frac{ 
{}_{
{}_{} 
}
{ \Big\langle  
( F_{1}( \Xi_1 ) (\Phi_{{1}  , {{2} } } 
F_2 ( \Xi_2)  ) ( \Phi^*_{{0}  , {{1} } } ( F_0 ( \Xi_0 )   \nu_0 )) ,
G_1( \Gamma_1 ) 
\Big\rangle  }
}
{ 
{
\Big\langle 
\nu_0,
F_0 ( \Xi_0 )  \Phi_{{0}  , {{1} } } \Big( F_{{1} } (\Xi_{{1} } ) \Phi_{{1}  , {{2} } }
\Big(   F_{{2} } (\Xi_{{2} } )    \Big) \Big) \Big\rangle  }
}
\\
&
\qquad
\qquad
\qquad
(\forall \Gamma_1 \in
{\cal G}_1
).
\tag{38}
\end{align*}

\par
\noindent
\bf 
{Example 3}
\rm
[Continued from the above example].
\rm
For each $t=1,2$,
assume that
$\Phi_{\pi(t), t}: C(\Omega_t) \to C(\Omega_{\pi(t)})$
is deterministic,
that is,
there exists a continuous map
$
\phi_{\pi(t), t}: \Omega_{\pi(t)} \to \Omega_t$
satisfying \textcolor{black}{(28)}.
And,
putting
$K=\{\delta_{\omega_0} \}$, 
consider the measurement
${\mathsf M}_{
C(\Omega_{0})
} ({}\widehat{\mathsf O}_0
\equiv ({\bigtimes}_{t \in T} X_t,$
${{\bigstimes}_{t \in T} {\cal F}_t}
,$
$ {\widehat F}_0), S_{[\ast]} {(\{ \delta_{\omega_0} \})}{})$.
Then,
we see,
by \textcolor{black}{(38)},
that,
for any 
$g_1$
in $C(\Omega_1)$,
\begin{align*}
&
\langle \nu_{\{1\}}^a , g_1
\rangle
\\
=
&
\frac
{ 
{ \Big\langle  \delta_{\omega_0},
F_0 ( \Xi_0 )  \Phi_{0  , 1} \Big( F_{1} (\Xi_{1} )  
g_1
\Phi_{1,2 }
\Big(   F_{{2} } (\Xi_{{2} } )   \Big) \Big) \Big\rangle  }
}
{ 
{
\Big\langle 
\delta_{\omega_0}
,
F_0 ( \Xi_0 )  \Phi_{{0}  , {{1} } } \Big( F_{{1} } (\Xi_{{1} } ) 
\Phi_{{1}  ,
 {{2} } }
\Big(   F_{{2} } (\Xi_{{2} } )   \Big) \Big) \Big\rangle  }
}
\\
=
&
\frac
{ 
{ 
[F_0 ( \Xi_0 )](\omega_0)
\big[ F_{1} (\Xi_{1} )  
g_1
\Phi_{1,2 }
\big(   F_{{2} } (\Xi_{{2} } )   \big) \big]
(\phi_{0,1}(\omega_0))
}
}
{ 
[F_0 ( \Xi_0 )](\omega_0)
\big[ F_{1} (\Xi_{1} )   \Phi_{1,2 }
\big(   F_{{2} } (\Xi_{{2} } )   \big) \big]
(\phi_{0,1}(\omega_0))
}
\\
=
&
g_1 (\phi_{0,1}(\omega_0))
=
\langle \delta_{\phi_{0,1}(\omega_0)} , g_1
\rangle.
\end{align*}
Thus,
we see that
\begin{align*}
\nu_{\{1\}}^a=\delta_{\phi_{0,1}(\omega_0)}.
\tag{39}
\end{align*}
Further we easily see that
$\nu_T^a$
$=$
$[B_{\widehat{\mathsf{O}}_{0} }(\bigtimes_{t \in T} \Xi_t)]
(\delta_{\omega_0})$
$=$
$\delta_{(\omega_0, \phi_{0,1}(\omega_0), \phi_{0,2}(\omega_0))}$
$(
\in {\cal M}_{+1}^p(
\Omega_0 \times
\Omega_1 \times
\Omega_2
)
)$.

\par
\par
\noindent
\subsection{\normalsize
Causal Fisher-Bayes Method
in Classical
$C(\Omega)$
}

\par
Now we can present
Corollary 4
(i.e.,
regression analysis in a wide sense)
as follows.
\begin{itemize}
\item[(W$_1$)]
$
{\text{[Corollary 4]}}
$
$=$
$\underset{\text{(Fisher's method)}}{
\text{
[Theorem 1]
}}
$
$+$
$\underset{\text{(Bayes' method)}}{
\text{
[Theorem 3]
}}
$
\end{itemize}

\rm

\par
\noindent
{\bf Corollary 4}
\rm
[Causal Fisher-Bayes method
(i.e.,
Regression analysis in a wide sense),
{\rm cf.$\;$}\textcolor{black}{\cite{Ishi7}}].

\it
Let
$t_0$
be the root
of a tree $T$.
Let
$[{\mathbb O}_T{}]$
$=$
$[{}
\{ {\mathsf{O}}_t ({}\equiv ({}X_t , {\cal F}_{t}, {F}_t ))
\}_{ t \in T} ,$
$
\{  \Phi_{t_1,t_2}{}: $
$C(\Omega_{t_2}) \to C(\Omega_{t_1}) \}_{(t_1,t_2) \in T^2_\le }$
$]$
be
a sequential causal observable
with
the realization
$\widehat{\mathsf{O}}_{t_0} $
$\equiv (\bigtimes_{t \in T } X_t , $
$\bigstimes_{t \in T } {\cal F}_{t} ,$
${\widehat F}_{t_0})$.
Assume the statistical measurement
${\mathsf M}_{
C(\Omega_{t_0})
}
(\widehat{\mathsf{O}}_{t_0}, S_{[\ast]}(K) )$.
And assume that
we know that
a measured value obtained by
the measurement
${\mathsf M}_{
C(\Omega_{t_0})
}
(\widehat{\mathsf{O}}_{t_0}, S_{[\ast]}(K) )$
belongs to
$\bigtimes_{t \in T} \Xi_t$.
Then,
there is a reason to infer that the mixed state
$\nu_T^{a} ( \in {\cal M}_{+1}^m(\bigtimes_{t \in T } \Omega_t ) )$
after the measurement
${\mathsf M}_{
C(\Omega_{t_0})
}
(\widehat{\mathsf{O}}_{t_0}, S_{[\ast]}(K) )$
is given by
$[B_{\widehat{\mathsf{O}}_{t_0} }(\bigtimes_{t \in T} \Xi_t)]
(\nu_0)
$.
Here,
the $\nu_0
(\in K)$
is defined by
\begin{align*}
&
\int_\Omega [{\widehat F}_{t_0}(\bigtimes_{t \in T} (\Xi_t)](\omega)\nu_0(d \omega )
\\
=
& 
\max_{\nu \in K } 
\int_\Omega 
[{\widehat F}_{t_0}(\bigtimes_{t \in T} (\Xi_t)]
(\omega)\nu(d \omega ).
\tag{40}
\end{align*}
\rm
\rm
\par
\vskip0.3cm
\par
\par
\noindent
{\bf Remark 10.}
Note that
Fisher maximum likelihood method
and
Bayes' theorem
are hidden in Corollary 4.
That is,
Corollary 4
includes the following procedure:
\begin{itemize}
\item[(W$_2$)]
$
\underset{(\subseteq {\cal M}_{+1}^m(\Omega_{t_0}))}{
K}
\xrightarrow[\text{\tiny Theorem 1}]{\text{\tiny Fisher}}
\underset{
(\in K)}
{\nu_0}
$
\\
$
\qquad
\qquad
\qquad
\qquad
\xrightarrow[\text{\tiny Theorem 3}]{\text{\tiny Bayes}}
\underset{(
\in {\cal M}_{+1}^m(\bigtimes_{t\in T} \Omega_t))}
{\nu_T^a}
$
\end{itemize}
which is the generalization of the \textcolor{black}{(K)}.
\par
\noindent
{\bf Answer 1}
\rm
[
Answer to \textcolor{black}{Problem 1 (S)}
].
Now we can answer Problem 1 \textcolor{black}{(S)} as follows.
The \textcolor{black}{(33)}
says that
$\nu_0$
$=$
$\delta_{(\alpha_0, \beta_0)}$
$=
\delta_{(0.4, 1.4)}$.
Thus,
using
\textcolor{black}{(39)},
we see that
$\nu_{\{1\}}^a$
$=$
$\delta_{\alpha_0 + \beta_0}$
$=
\delta_{1.8}$.
Also, note that
\textcolor{black}{(33)}
and
\textcolor{black}{(39)}
are consequences of Corollary 4.
Hence,
the calculation \textcolor{black}{(34)}
is justified
by
Corollary 4.


%
\par
\par
\noindent
\section{
\large
Conclusions
}
\par
\noindent


%
It is a matter of course that our problem 
(A$_3$)
(i.e.,
What is statistics?)
is most fundamental in sceince.
Thus,
there is every reason to consider that,
in order to solve this problem,
we have to
start from {\lq\lq}worldview".
Hence,
in this paper
we started from the linguistic worldview.
\par
\noindent
\par
\noindent
\par
Most scientists may be skeptical about traditional and modern philosophy
unless they know the linuistic world view
(A$_2$) in {\bf Figure 1}.
Thus, we believe that
{\bf Figure 1}
(particularly, 
\textcircled{\scriptsize 3}
and
\textcircled{\scriptsize 4}
)
implies and realizes the end of grand narratives
(i.e.,
the 3000 years' final answer to the worldview problem).
If it be so,
we can assert that
our proposal in this paper
is decisive and final.
That is, we assert that
\begin{itemize}
\item[(X)]
to do science (other than physics) is to describe every phenomenon by
quantum langugae,
\end{itemize}
which may be also regarded as the answer to the question:
"What is science?"

We hope that 
our proposal
(i.e.,
the quantum linguistic formulation of statistics)
will be examined 
from various points of view.

%
%
\rm
\vskip-0.5cm
\par
\renewcommand{\refname}{
\large 
References}
{
\small

\normalsize
}

%

\par
\noindent


\begin{thebibliography}{9}
\rm
\bibitem{Ishi1}
S. Ishikawa,
\newblock {\rm {\lq\lq}Fuzzy Inferences by Algebraic Method,{\rq\rq}}
\newblock {\em Fuzzy Sets and Systems},
{Vol. 87}, No. 2, 1997, pp.181-200.
\\
\href{http://dx.doi.org/10.1016/S0165-0114(96)00035-8}{doi: 10.1016/S0165-0114(96)00035-8}
\\
\footnotesize{
(\url{http://www.sciencedirect.com/science/article/pii/S0165011496000358
}
)
}
%
%
%
\bibitem{Ishi2}
S. Ishikawa,
{\rm {\lq\lq}A Quantum Mechanical Approach to Fuzzy Theory,{\rq\rq}}
{{\em Fuzzy Sets and Systems}}, 
{Vol. 90}, No. 3, 1997, pp. 277-306.
\\
\href{http://dx.doi.org/10.1016/S0165-0114(96)00114-5}{doi: 10.1016/S0165-0114(96)00114-5}
\\
\footnotesize{
(\url{http://www.sciencedirect.com/science/article/pii/S0165011496001145
}
)
}
%
%
\bibitem{Ishi3}
S. Ishikawa,
{\rm {\lq\lq}Statistics in measurements},"
{\it Fuzzy sets and systems}, 
{Vol. 116}, No. 2, 141-154 (2000).
\\
\href{http://dx.doi.org/10.1016/S0165-0114(98)00280-2}{doi:10.1016/S0165-0114(98)00280-2}
\\
\footnotesize{
(\url{http://www.sciencedirect.com/science/article/pii/S0165011498002802
}
)
}
%
%
%
\bibitem{Ishi4}
S. Ishikawa,
{\rm {\lq\lq}Mathematical Foundations of Measurement Theory,{\rq\rq}}
Keio University Press Inc. 335pages,
2006.
\\
(\url{http://www.keio-up.co.jp/kup/mfomt/})
\bibitem{Ishi5} {S. Ishikawa,}
\newblock {\rm {\lq\lq}A New Interpretation of Quantum Mechanics,{\rq\rq}}
\newblock {\em Journal of quantum information science},
{Vol. 1}, No. 2, 2011, pp.35-42.
\\
\href{http://dx.doi.org/10.4236/jqis.2011.12005}{doi: 10.4236/jqis.2011.12005}
\\
\url{http://dx.doi.org/10.4236/jqis.2011.12005}
\bibitem{Ishi6} {S. Ishikawa,}
\newblock {\rm {\lq\lq}
Quantum Mechanics and the Philosophy of Language:
Reconsideration of Traditional
Philosophies,{\rq\rq}}
\newblock {\em Journal of quantum information science},
{Vol. 2}, No. 1, 2012, pp.2-9.
\\
\href{http://dx.doi.org/ 10.4236/jqis.2012.21002}{doi: 10.4236/jqis.2012.21002}
\\
\url{http://dx.doi.org/ 10.4236/jqis.2012.21002}
\bibitem{Ishi7} {S. Ishikawa,}
\newblock {\rm {\lq\lq}A Measurement Theoretical
Foundation of Statistics,{\rq\rq}}
\newblock {\em Applied Mathematics},
{Vol. 3}, No. 3, 2012,
pp. 283-292.
\\
\href{http://dx.doi.org/10.4236/am.2012.33044}{doi: 10.4236/am.2012.33044}
\\
\url{http://dx.doi.org/10.4236/am.2012.33044}
\bibitem{Ishi8} S. Ishikawa,
\newblock {{\lq\lq}The Linguistic Interpretation of Quantum Mechanics,{\rq\rq}}
\href{http://arxiv.org/abs/1204.3892v1}{arXiv:1204.3892v1}[physics.hist-ph],
\newblock {2012.}
\\
\url{http://arxiv.org/abs/1204.3892v1}
\bibitem{Ishi9} {S. Ishikawa,}
\newblock {\rm {\lq\lq}Ergodic Hypothesis and Equilibrium Statistical 
Mechanics in the Quantum Mechanical World View,{\rq\rq}}
\newblock {\em World Journal of Mechanics},
{Vol. 2}, No. 2, 2012,
pp. 125-130.
%
\\
\href{http://dx.doi.org/10.4236/wjm.2012.22014}{doi: 10.4236/wjm.2012.22014}
\\
\url{http://dx.doi.org/10.4236/wjm.2012.22014}
\newpage
\bibitem{Ishi10} S. Ishikawa,
\newblock {{\lq\lq}Zeno's paradoxes in the Mechanical World View,{\rq\rq}}
\href{http://arxiv.org/abs/1205.1290v1}{arXiv:1204.3892v1}[physics.hist-ph],
\newblock {2012.}
\\
\url{http://arxiv.org/abs/1205.1290v1}
\bibitem{Ishi11} {S. Ishikawa,}
\newblock {\rm {\lq\lq}Monty Hall Problem and the Principle of Equal 
Probability
in Measurement Theory,{\rq\rq}}
\newblock {\em Applied Mathematics},
{Vol. 3}, No. 7, 2012,
(to appear).
\bibitem{IshiU} S. Ishikawa,
\newblock {\rm {\lq\lq}Uncertainty relation
in simultaneous measurements for arbitrary observables,"}
\newblock {\it Rep. Math. Phys.,}
\newblock {\rm {Vol. 29, No. 3}},
\newblock {pp. 257-273},
\newblock {1991}
\\
\href{http://dx.doi.org/10.1016/0034-4877(91)90046-P}{doi: 10.1016/0034-4877(91)90046-P}
\\
\footnotesize{
(\url{http://www.sciencedirect.com/science/article/pii/003448779190046P
}
)
}
\bibitem{Neum} {J. von Neumann,}
\newblock {\rm {\lq\lq}Mathematical Foundations of Quantum Mechanics,{\rq\rq}}
\newblock {\rm Springer Verlag, Berlin,}
\newblock {1932.}
\bibitem{Saka}
S. Sakai,
{\it $C^*$-algebras and $W^*$-algebras}, 
Ergebnisse der Mathematik und ihrer Grenzgebiete (Band 60), 
Springer-Verlag, (1971)
\bibitem{Yosi}
K. Yosida,
{\rm {\lq\lq}Functional Analysis,
{\rq\rq}}
Springer-Verlag, 6th edition, 1980.
\bibitem{Davi} E. B. Davies,
\newblock {\rm {\lq\lq}Quantum Theory of Open Systems,{\rq\rq}}
\newblock {Academic Press,}
\newblock {1976.}
%
\bibitem{Kolm}
A. Kolmogorov,
{{\lq\lq}Foundations of the Theory of Probability (Translation),{\rq\rq}}
Chelsea Pub Co. Second Edition,
New York,
1960,
\end{thebibliography}
\end{document}